\documentclass{aa}  
\usepackage{graphicx}
\usepackage{amsmath}    
\usepackage{amssymb}    
\usepackage{relsize}
\usepackage{txfonts}
\usepackage[colorlinks=true,linkcolor=blue,allcolors=blue]{hyperref}%

\newcommand{\Msun}{M$_\odot$}

\newcommand{\Mstar}{\ensuremath{M_*}}
\newcommand{\rmaxhot}{\ensuremath{r|_{{\rm max}(p_{\rm hot})}}}
\newcommand{\rcut}{\ensuremath{r_\mathrm{cut}}}
\newcommand{\rmax}{\ensuremath{r_\mathrm{max}}}
\newcommand{\Mshalo}{\ensuremath{M_{*,\mathrm{halo(r< 2R_e)}}}}

\begin{document} 

   \title{Mass of the dynamically hot inner stellar halo predicts the ancient accreted stellar mass}

   \subtitle{}

   \author{Ling Zhu$^1$\thanks{E-mail: lzhu@shao.ac.cn}, Annalisa Pillepich$^2$, Glenn van de Ven$^{3}$, Ryan Leaman$^{2,3}$,  Lars Hernquist$^{4}$,  Dylan Nelson$^{5}$, Ruediger Pakmor$^{6}$,  Mark Vogelsberger$^{7}$, Le Zhang$^{1}$
   }


   \institute{
Shanghai Astronomical Observatory, Chinese Academy of Sciences, 80 Nandan Road, Shanghai 200030, China\\
\email{lzhu@shao.ac.cn}
\and
Max Planck Institute for Astronomy, K\"onigstuhl 17, 69117 Heidelberg, Germany 
\and
Department of Astrophysics, University of Vienna, T\"urkenschanzstra{\ss}e 17, 1180 Vienna, Austria
\and
Harvard--Smithsonian Center for Astrophysics, 60 Garden Street, Cambridge, MA 02138  
\and
Universit\"{a}t Heidelberg, Zentrum f\"{u}r Astronomie, Institut f\"{u}r theoretische Astrophysik, Albert-Ueberle-Str. 2, 69120 Heidelberg, Germany. 
\and
Max-Planck-Institut f{\"u}r Astrophysik, Karl-Schwarzschild-Str. 1, 85741 Garching, Germany 
\and
Kavli Institute for Astrophysics and Space Research, Massachusetts Institute of Technology, Cambridge, MA 02139, USA 
             }

   \date{Received; accepted}
   
   \titlerunning{The information content of the dynamically hot inner stellar halo}
\authorrunning{Zhu et al.}  
 
  \abstract
   {Galactic dynamical structures are fossil records of the assembly histories of galaxies. By analyzing the cosmological hydrodynamical simulation TNG50, we find that a dynamical structure that we call the ``hot inner stellar halo,'' defined by stars on dynamically hot orbits with circularity $\lambda_z < 0.5$ at $3.5\,{\rm kpc}<r \lesssim 2\,R_e$, is a strong indicator of the mass of accreted satellite galaxies. 
    We find a strong correlation between the mass of this hot inner stellar halo and the total ex situ stellar mass. There is a similarly strong correlation with the stellar mass of the most massive secondary galaxy ever merged. 
   These TNG50 correlations are compatible with those predicted by other simulations, for example by TNG100 across the whole mass range under study (galaxy stellar masses, $M_*$, in the $10^{10.3-11.6}$\,\Msun\,  range) and by EAGLE for $M_* \gtrsim 10^{10.6} $\,\Msun\, galaxies.\ This shows that our predictions are robust across different galaxy formation and feedback models and hold across a wide range of numerical resolution. The hot inner stellar halo is a product of massive and typically ancient mergers,  with inner-halo stars exhibiting three main physical origins: accreted and stripped from massive satellites, dynamically heated by mergers from the bulge and/or disk in the main progenitor, and formed from star formation triggered during mergers. The mass of the hot inner stellar halo defined in this paper is a quantity that can be robustly obtained for real galaxies by applying a population-orbit superposition method to integral-field-unit spectroscopy data, out to a distance of $\sim2\,R_e$, which is possible with current observations. Hence, this paper shows that integral-field-unit observations and dynamical models of the inner regions of galaxies provide a way to quantitatively determine the mass of ancient accreted satellites.  
   }

\keywords{galaxies: formation -- galaxies: evolution -- galaxies: simulations -- galaxies: stellar kinematics}

\maketitle


\section{Introduction}
Galaxies can experience a diversity of assembly and merger histories, which in turn are thought to lead to the observed variety in galaxy star formation rates, stellar morphologies, and kinematics. Galaxies with quiescent merger histories are thought to be likely dominated by fast rotating, flat stellar disks with recent star formation \citep[e.g.,][]{Fall1980}. On the other hand, major (non-gas-rich) mergers can play a key role in destroying stellar disks and in creating dynamically hot spheroidal components in the remnant galaxies \citep{Cox2006,Hoffman2010,Bois2010, Bois2011, Naab2014, Pillepich2015}. Minor mergers are thought to be efficient in building the outer stellar halos \citep{Naab2009}.

With deep photometric imaging, observed galaxies can be decomposed into multiple stellar components, such as a disk, bulge, bar, and halo. Great efforts over the past few years have been made to find the connection between galactic substructures and merger events. For example, it has been suggested that the bulges of Milky Way-mass galaxies have not been built up by mergers \citep{Kormendy2004}, or at least not exclusively \citep{Bell2017}.
On the other hand, deep imaging does not yet seem to provide enough information to infer the details of the accreted satellites that formed the stellar halos \citep{Spavone2020}, and their very separation from other galactic components may be affected by large uncertainties in comparison to kinematically based decomposition methods \citep{Du2020}. 

Over the last two decades, the rapidly growing field of integral-field-unit (IFU) spectroscopy has significantly deepened our knowledge of stellar structures in galaxies across a wide range of masses and Hubble types. 
A number of IFU surveys, such as SAURON \citep{Davies2001}, ATLAS-3D \citep{Cappellari2011}, CALIFA \citep{Walcher2014}, SAMI \citep{Croom2012}, and MaNGA \citep{Bundy2015}, have provided kinematic maps as well as stellar age and metallicity maps for thousands of nearby galaxies. By applying an orbit-superposition dynamical model to the IFU data \citep{Cappellari2007, vdB2008, Zhu2018a}, it has been possible to infer the internal orbital structure of individual galaxies within the data coverage. Two parameters are usually used to characterize the orbits in these models: the circularity, $\lambda_z$, which represents the angular momentum of the stars, and the {\rm time-averaged radius}, $r$, as a proxy for binding energy. The distribution of stellar orbits in a galaxy can thus be represented by the probability density of orbits in the phase-space of circularity, $\lambda_z$, versus radius, $r$. The multiple structures of galaxies can also be identified within these phase-space planes with, for example, the subdivision of the stellar component in a galaxy into cold ($\lambda_z > 0.8$), warm ($0.2<\lambda_z<0.8$), hot ($|\lambda_z|<0.2$), and counter-rotating ($\lambda_z < -0.2$) orbits.

In recent years, the stellar orbit circularity distribution within $1\,R_e$ has been obtained for 300 CALIFA galaxies across the Hubble sequence \citep{zhu2018b} and for a few hundred early-type galaxies in MaNGA \citep{Jin2020}. These works show that the fraction of dynamically hot orbits increases from low-mass to high-mass galaxies, consistently with an increase in the mass fraction of the traditional bulge \citep{Weinzirl2009}. The IFU-based orbital decomposition of large numbers of galaxies is now achievable \citep{zhu2018c}. However, the formation of these orbital components has not yet been quantitatively connected to galaxies' merger histories.

The details of merger histories have been (partially) uncovered in the closest galaxies, where single stars in the stellar halos can be resolved, for example in the case of the Milky Way \citep{Helmi2018, Belokurov2018, Helmi2020}, the Andromeda galaxy \citep{DSouza2018, DSouza2018b}, and NGC 5128 \citep{Rejkuba2011}, or where individual globular clusters in the halos are detected \citep{Forbes2016, Beasley2018,Kruijssen2019}, or where streams have been detected \citep{Merritt2016, Harmsen2017}. Recently, studies have attempted to constrain the global ex situ fractions \citep{Davison2021, Boecker2020} or the mass of satellite mergers \citep{Pinna2019,Martig2021} with stellar population distributions obtained from IFU data in some edge-on lenticulars \citep{Poci2019,Poci2021}. 
However, so far the assembly and merger histories of most galaxies remain hidden, including in the nearby Universe.

From a numerical perspective, recent and current cosmological hydrodynamical simulation projects \citep{Vogelsberger2020}, such as IllustrisTNG\footnote{\url{www.tng-project.org}} and EAGLE\footnote{\url{http://eagle.strw.leidenuniv.nl/}}, have been able to reproduce large numbers of galaxies across the morphological spectrum with well-resolved structures \citep{Pillepich2019, Pulsoni2020, Du2019, Correa2017}. For example, the TNG100 and TNG300 simulations of the IllustrisTNG project produce galaxy morphological structures, such as galaxy concentration and bulge strength, that  statistically match observational constraints for galaxies with $M_*\gtrsim 10^{9.5}$\,\Msun\, at low redshift \citep{Rodriguez-Gomez2019}, greatly improving upon the analog outcome of the original Illustris simulation\footnote{\url{www.illustris-project.org}}. The stellar sizes of TNG100 and TNG300 galaxies have been shown to be in good agreement with observations at $10^9\lesssim M_*\lesssim 10^{11}$\,\Msun\,, to within 0.1-0.2 dex across the full mass range \citep{Genel2018}. With the higher resolution of TNG50, galactic structures, in particular the stellar disk scale height and the stellar size of low-mass objects \citep{Pillepich2019}, match the observations even better \citep{Zanisi2021}. Similarly, the EAGLE simulation has been shown to successfully produce galactic structures, such as bulge sizes, that are similar to observations in galaxies with $M_*\gtrsim 10^{10}$\,\Msun\, \citep{Lange2016}; it has also produced galaxy stellar sizes that match observations well in galaxies  with $10^{10}\lesssim M_* \lesssim 10^{11}$\,\Msun\  and are about 0.2 dex larger for lower or higher galaxy masses \citep{Wang2020}, at least when comparisons are done at face value (see \citealt{deGraaff2021} for a discussion).   
  
With the simulation data, and thanks to the realism of their simulated galactic structures, it has also been possible to trace back the formation of galactic structures in great detail \citep{Lagos2018, Obreja2019,Trayford2019,Du2021, Pulsoni2021}, including adopting dynamical decomposition methods similar to those applied to observational data. For example, by choosing the same definition of cold, warm, and hot components, \cite{Xu2019} systematically compared the orbital structures of TNG100-simulated galaxies to CALIFA galaxies and showed that the orbital fractions as a function of stellar mass in TNG100 are quantitatively consistent with the observed ones.

The availability of flexible dynamical decomposition methods that are applicable to large numbers of observed galaxies and the realism of current cosmological galaxy simulations means it is now possible to quantitatively combine the two to address the question of whether we can identify galactic structures in present-day galaxies that (1) are physically connected to their past merger events and (2) can be directly compared across simulation outputs and observations, given the unavoidable limitations on both ends. In this paper we specifically address this goal by focusing on the hot inner stellar halos of galaxies defined by stars with $\lambda_z<0.5$ and $3.5\mathrm{kpc}<r<2R_e$. 

In the Milky Way, stars on highly radial orbits in the vicinity of the Sun have been associated with the Galaxy's inner stellar halo and have been used as a probe of its ancient massive merger event(s) -- such as the so-called Gaia Enceladus or Gaia Sausage progenitor \citep{Helmi2018, Belokurov2018}. Such structures in Milky Way-like galaxies can also arise in galaxy formation simulations \citep[e.g.,][]{Grand2020}. In this paper we address whether dynamical structures that are similar to the inner stellar halo of the Galaxy and that directly trace the ancient mergers generally exist across the mass and galaxy-type spectrum and whether they can be detected in external galaxies with current IFU data.

\begin{figure*}
\includegraphics[width=18cm]{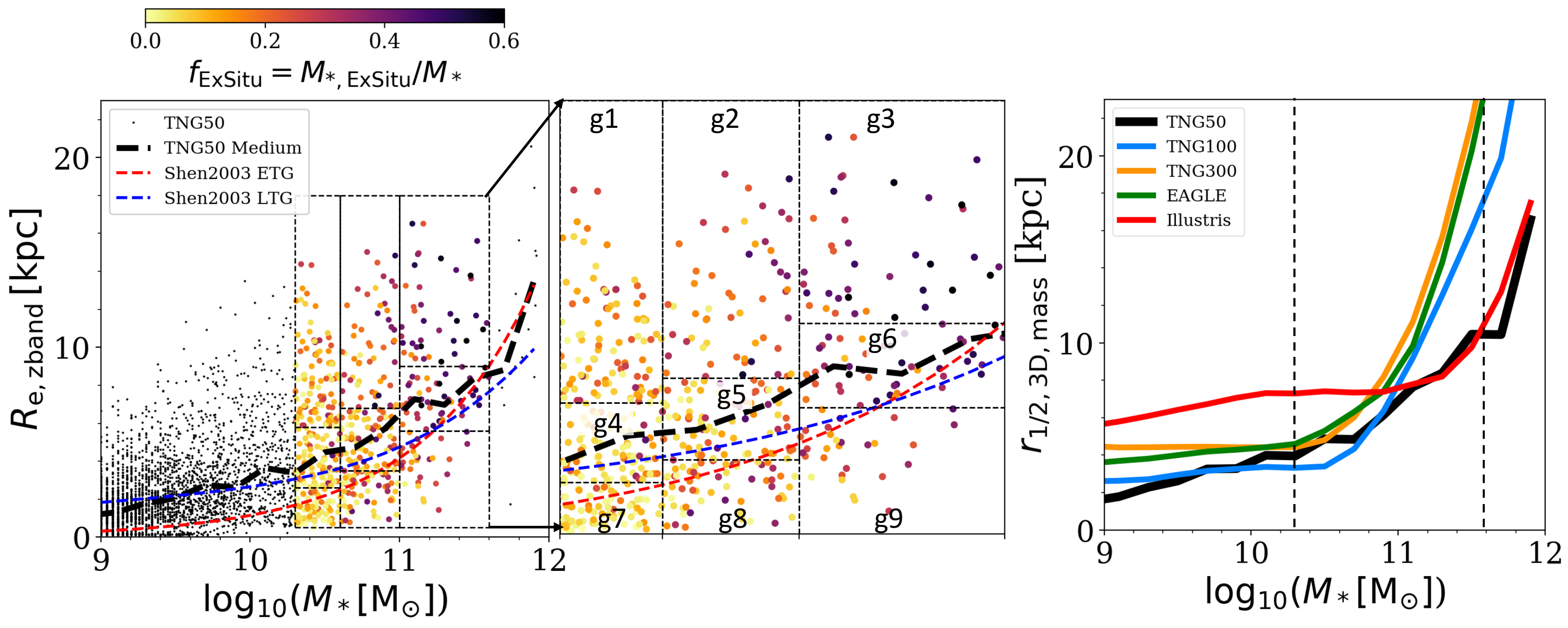}
\caption{{Stellar mass and size of the simulated galaxies analyzed in this paper}. {\it Left and central panels:} Total stellar mass, \Mstar, versus projected half-light size in the z band, $R_{\rm e,zband}$, of TNG50 galaxies at $z=0$. The dashed black curve is the median size of TNG50 galaxies as a function of \Mstar. The red and blue curves show the median sizes of early-type and late-type SDSS galaxies \citep{Shen2003} for reference. 
TNG50 galaxies with $10^{10.3}<M_*<10^{11.6}\,M_{\odot}$ are color coded by their stellar ex situ fraction (see also the zoomed-in version in the middle panel). The ex situ fraction systematically increases with increasing stellar mass and size. However, there are still large variations among galaxies of similar mass and size. We further divide the TNG50 samples into nine subsamples as indicated by the dashed squares labeled g1-g9 from left to right and from top to bottom. {\it Right panel:} Median stellar sizes as a function of \Mstar\, for TNG50, TNG100, TNG300, EAGLE, and Illustris galaxies at $z=0$. Here we use the 3D half-mass size, $r_{\rm 1/2,3D, mass}$, for all galaxies. The vertical dashed lines mark the mass regions we focus on in this paper. 
}
\label{fig:MRe}
\end{figure*}

In practice, in this paper we analyze thousands of galaxies from the IllustrisTNG and EAGLE simulations, identify their hot inner stellar halo components in a way that is also consistent and achievable with current IFU observations, and -- via the information provided by the merger trees of the simulations -- look for quantitative connections between the inner stellar-halo mass and properties of the merger and stellar assembly histories. In particular, we focus on results from the TNG50 simulation and, after comparing them to those from other IllustrisTNG and EAGLE runs, use it to identify the main formation channels of the hot inner stellar halos of simulated galaxies. Our analysis is tailored toward the information that is obtainable from the Fornax3D survey \citep{Sarzi2018}, which uses MUSE/VLT and is designed to cover at least the inner $2\,R_e$ of galaxies. In a companion paper, we will show that the stellar orbit distribution, as well as the stellar age and metallicity distributions, of each Fornax3D galaxy can be obtained through the population-orbit superposition method of \cite{Zhu2020}, hence significantly improving upon the orbit characterization within $1\,R_e$ with CALIFA or MaNGA observations and at last starting to probe into the stellar halos.  

The paper is organized as follows: in Sect. 2 we describe the adopted cosmological simulations; in Sect. 3 we investigate the orbital decomposition of the simulated galaxies and define their hot inner stellar halos; in Sect. 4 we present the main results of the paper, which are then discussed and summarized in Sects. 5 and 6, respectively.

\section{Cosmological galaxy simulations}
\label{simulations}
\begin{table}
\def\arraystretch{1.5}
\caption{Basic information regarding the publicly available cosmological simulations we use or refer to in this paper.}
\scriptsize\centering
\label{tab:sim_info}
\begin{tabular}{*{5}{l}}
\hline
       Name  & Softening length [kpc] & $m_{\rm baryon}$ [$M_{\odot}$]  & $m_{\rm DM}$ [$M_{\odot}$] & L$_{\rm box}$ [cMpc]\\
       \hline
TNG50 & 0.3 &  8.5e4 & 4.5e5 & 50\\
TNG100 & 0.7 &  1.4e6 & 7.5e6 & 100\\
TNG300 & 1.5 &  1.1e7 & 5.9e7 & 300\\
EAGLE & 0.7 &  1.8e6 & 9.7e6 & 100\\
Illustris & 0.7 &  1.3e6 & 6.3e6 & 100\\
\hline
\hline
\end{tabular}
\tablefoot{From left to right, the columns show the name of the simulation, the gravitational softening length, mass of stellar particles or gas cells, mass of dark matter (DM) particles, and the side length of the simulation box. All values are approximate.}
\end{table}

\subsection{TNG50 and IllustrisTNG}
\label{sec:TNG}
Among the existing cosmological hydrodynamical simulations for the formation and evolution of galaxies, the IllustrisTNG simulations \citep[hereafter TNG;][]{Springel2018, Marinacci2018, Naiman2018, Pillepich2018b, Nelson2018} have been particularly successful in reproducing a broad range of observational findings \citep{Nelson2019release}. These include the galaxy mass-size relation at $0<z<2$ already mentioned in the Introduction \citep{Genel2018}, but also the gaseous and stellar disk sizes and heights \citep{ Pillepich2019}, galaxy colors, the stellar age and metallicity trends at $z\sim0$ as a function of galaxy stellar mass in comparison to SDSS results \citep{Nelson2018}, and resolved star formation in star-forming galaxies \citep{ENelson2021}, as well as the characteristics of the stellar orbit distributions from the CALIFA survey \citep{Xu2019} and the kinematics of early-type galaxies in comparison to data from ATLAS-3D, MaNGA, and SAMI \citep{Pulsoni2020}. 

All the details of the numerical and physical model implemented in the TNG simulations are given in the method papers by \citet{Weinberger2017} and \citet{Pillepich2018a}. 
Furthermore, the suite comprises three flagship runs: TNG50, TNG100, and TNG300, with a cosmological volume of $(50\,\rm Mpc)^3$, $(100\,\rm Mpc)^3$, and $(300\,\rm Mpc)^3$ and a stellar particle resolution of $8.5 \times 10^{4}$\,\Msun, $1.4 \times 10^{6}$\,\Msun, and $1.1 \times 10^{7}$\,\Msun, respectively \citep{Nelson2019release}. The basic information of these simulations is given in Table~\ref{tab:sim_info}.

For most of the analysis in this paper, we use data from the TNG50 run \citep{Pillepich2019, Nelson2019}. It is uniquely well suited for our study as (1) it reaches the numerical resolution typical of many zoomed-in simulation projects -- with stellar particles of $8.5 \times 10^{4}$\,\Msun, so that the inner structural details of galaxies, such as thin disks, thick disks, bulges, and halos, are fairly well resolved --
and (2) it still offers a large sample of galaxies, including massive ones, in a representative cosmological volume that encompasses diverse environments. 

In the TNG simulations, halos and subhalos, and thus the basic properties of galaxies, are identified using the friends-of-friends (FoF) and \textsc{Subfind} algorithms \citep{Springel2001, Dolag2009}. 
Based on \textsc{Subfind}, there are 548, 4002, and 50291 galaxies in the TNG50, TNG100, and TNG300 boxes at $z=0$, respectively, with $10^{10.3}<M_*<10^{11.6}$\,\Msun.

The histories of all galaxies in the simulations can be followed via the merger trees, as constructed by the \textsc{Sublink-gal} code based on the baryonic component of subhalos \citep{Rodriguez2015}. In this algorithm, each galaxy is assigned a unique descendant. We can identify for each galaxy its ``mergers'' -- secondary galaxies with a well-defined ``infall'' time that merge with the primary (i.e., do not exist after ``coalescence'' as individually identified \textsc{Subfind} objects). For each galaxy at a given time we can also identify its ``satellites'' -- galaxies with a well-defined infall time that orbit or fly by the primary prior to their possible ultimate merging.

Thanks to the merger trees, we can also distinguish between in situ and ex situ (i.e., accreted) stars \citep{Rodriguez-Gomez2019, Pillepich2018a}. For any galaxy at $z=0$, we call its stars ``in situ'' if they formed in a progenitor that belongs to the
main-progenitor branch of the galaxy, and ``ex situ'' otherwise. Ex situ stars are hence stellar particles stripped and consequently accreted into a galaxy: the ex situ stars of a galaxy are mostly accreted from its mergers (i.e., from objects that are already merged at the time of inspection); however, a fraction of ex situ stars can also arise from the stripping of orbiting and not-yet destroyed satellites.

\subsection{EAGLE}
The EAGLE \citep{Schaye2015, Crain2015} simulations have also been shown to successfully reproduce a range of observations of galactic properties, including the galaxy stellar mass function, the Tully-Fisher relation, and the galaxy mass-size relation. As mentioned in the previous section, the galaxy sizes as a function of stellar mass generally agree with the SDSS results, with differences smaller than 0.2 dex at $10^{10}\lesssim M_*\lesssim 10^{11}$\,\Msun. Galactic structures, such as disks and bulges, are well resolved, and the Hubble sequence is in place \citep{Lange2016, Trayford2019}. 

In this work we use publicly available data \citep{McAlpine2016} of the fiducial EAGLE simulation with a cosmological volume of about $(100\,\rm Mpc)^3$, with a stellar particle mass of $\sim 10^6$\,\Msun\, and a  Plummer equivalent gravitational softening of 0.7 in proper kpc at redshift $z=0$. The galaxy formation model and numerical code that produced the EAGLE outcome differ substantially from those of TNG. Here, we work with EAGLE halos and subhalos as identified by the same \textsc{FoF} group finder and \textsc{Subfind} algorithms adopted for the TNG simulations, based on which we also ran the same merger tree code as for TNG \citep{Nelson2019release}. The analysis of the EAGLE and TNG simulations is thus identical in methodology, including the merger and assembly definitions. There are 2049 galaxies in the EAGLE box with $10^{10.3}<M_*<10^{11.6}$\,\Msun\ at $z=0$.

\subsection{TNG and EAGLE galaxies adopted in this work}
\label{s:selection}
We use TNG50 for all the analysis throughout the paper. TNG100, TNG300, and EAGLE are only used to check if our results are robust against different galaxy formation and feedback models and across a range of numerical resolution. 

We selected simulated galaxies based on their galaxy stellar mass and stellar size. The left panel of Fig.~\ref{fig:MRe} shows the stellar size-mass plane for TNG50 at $z=0$. We define galaxy stellar mass as the total mass of all stars that are gravitationally bound to a galaxy according to the \textsc{Subfind} algorithm. For TNG50 galaxies, we calculated the circularized projected half-light sizes in the $z$ band, $R_{\rm e,zband}$, to make them comparable with results from SDSS galaxies \citep{Shen2003}. Most galaxies are late-type at $\Mstar\sim10^{10}$\,\Msun\, and are early-type at $\Mstar>10^{11}$\,\Msun. The sizes of TNG50 galaxies are generally consistent with the stellar size-mass relation of SDSS galaxies at $\Mstar>10^{10}$\,\Msun, as already quantified in depth by, for example, \cite{Zanisi2021}.

The TNG50 galaxies of Fig.~\ref{fig:MRe} are also color coded by their stellar ex situ fraction ($f_{\rm ExSitu} = M_{\rm *, ExSitu}/M_*$) following the definition and methods used earlier \citep{Rodriguez2016, Pillepich2018b}. The stellar ex situ fraction of galaxies increases with increasing stellar mass and size, but still with a large scatter in galaxies of similar mass and size. This result confirms a similar trend predicted by the EAGLE model \citep{Davison2020}. 

Throughout the paper, we focus on galaxies with stellar masses in  the range $10^{10.3}-10^{11.6}$ \Msun\,, irrespective of whether they are centrals or satellites. We further divide the TNG50 samples into nine subsamples.
We first made three bins of stellar mass, then three bins in size for each mass bin.
The sizes are divided as $R_e > \overline{R_e}+\sigma(R_e)/2$, $\overline{R_e}-\sigma(R_e)/2<R_e < \overline{R_e}+\sigma(R_e)/2$, and $R_e < \overline{R_e}-\sigma(R_e)/2$, and thus the numbers of galaxies in the three size bins are similar. Here $\overline{R_e}$ is the median and $\sigma(R_e)$ is the $1\sigma$ scatter of $R_e$ in each mass bin. The nine subsamples are labeled as g1-g9 from left to right and from top to bottom. 
In the right panel of Fig.~\ref{fig:MRe}, we compare the galaxy size-mass relations for TNG50, TNG100, TNG300, and EAGLE. 
Here we simply use the 3D half-mass radius, $r_{\rm 1/2, 3D, mass}$.
TNG50, TNG100, TNG300 and EAGLE galaxies have similar sizes at $\Mstar<10^{11}$\,\Mstar\, and match the observations well, whereas TNG100, TNG300, and EAGLE galaxies have sizes about 0.2 dex larger than observations at the high-mass end \citep{Genel2018}. 

The sizes of Illustris galaxies \citep{Vogelsberger2014, Vogelsberger2014b, Genel2015, Nelson2015} are also shown for comparison (red curve). As noted in previous papers \citep{Pillepich2018a, Rodriguez-Gomez2019}, Illustris galaxies have larger sizes at $\Mstar<10^{11}$\,\Mstar\, in comparison to the other simulations and to observations: in fact, an improvement on galaxy sizes was one of the goals of the TNG project \citep{Pillepich2018a}. As realistic galactic structures are a key requirement for us to define the hot inner stellar halo in a coherent way \citep{Rodriguez-Gomez2019}, we do not not use the original Illustris simulation to extract the main results of this paper. 

\begin{figure*}
\centering\includegraphics[width=16cm]{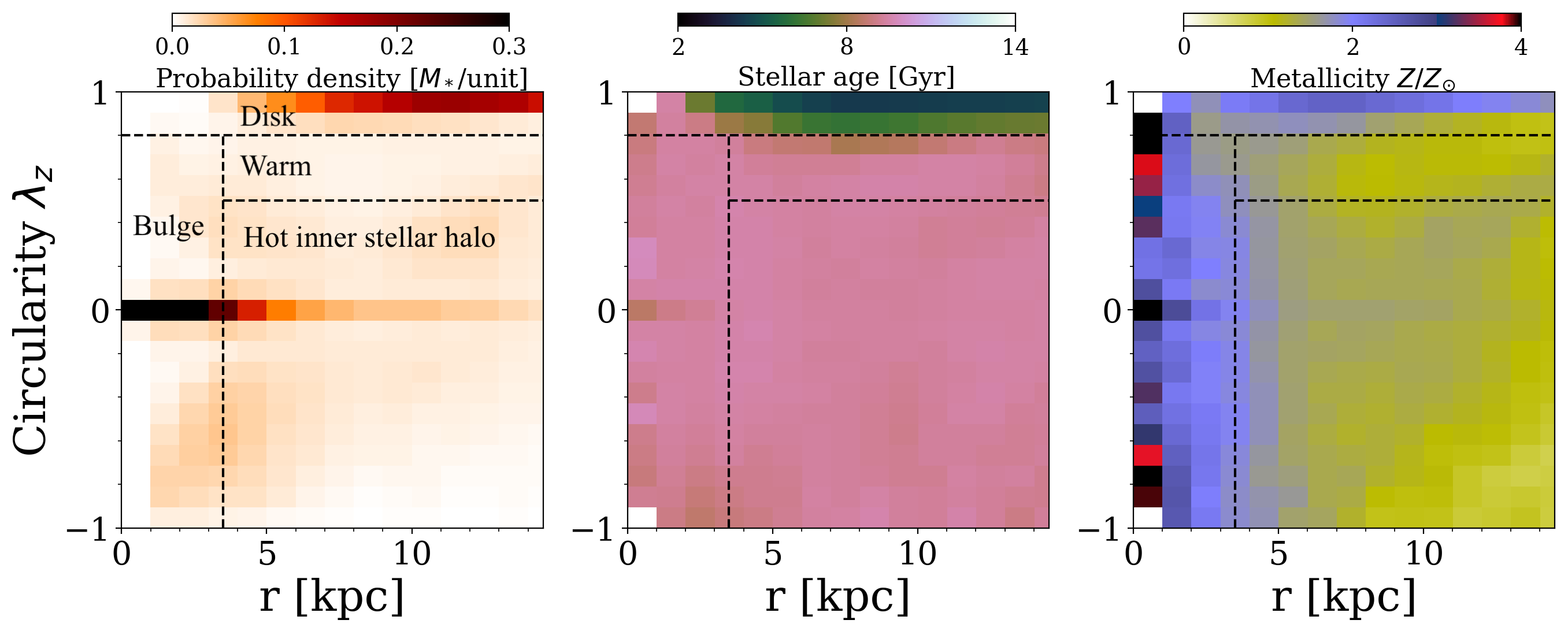}
\caption{{Definition of hot inner stellar halo advocated for and adopted in this paper.} We show the stellar orbit distribution, $p(r, \lambda_z)$, stellar age distribution, $t(r, \lambda_z)$, and metallicity distribution, $Z(r, \lambda_z)$, in the phase-space of circularity, $\lambda_z$, versus radius, $r$, of a simulated galaxy, TNG50 468590. We define as a hot inner stellar halo the galactic component whose stars have orbits with $r_{\rm cut}<r<r_{\rm max}$ and $\lambda_z < \lambda_{z,{\rm cut}}$, where $r_{\rm cut}$, $r_{\rm max}$, and $\lambda_{z,{\rm cut}}$ are, respectively, 3.5 kpc, $\max(7\,\rm{kpc}, 2R_e)$, and 0.5.}
\label{fig:rlz_468590}
\end{figure*}

\section{Dynamical decomposition of galaxies}

We aim to separate the stellar constituents of the simulated galaxies into kinematically motivated ``morphological'' components, such as disk, bulge, and hot inner stellar halo, to then focus on the formation channel of the hot inner stellar halo and its information content.

\subsection{Orbital properties}
We characterize the stellar orbits of simulated galaxies with two parameters, the {\rm time-averaged} radius, $r$, and circularity, $\lambda_z$. We measured these values with two complementary approaches.

In the first method, the radius and circularity of each stellar particle were calculated as the averages of the corresponding quantities for particles stored with equal time steps along the orbits. For the simulated galaxies, we have the instantaneous values of the stellar particles at $z=0,$ but we can integrate the orbits within the final potential. Namely, we froze the stellar particles of a simulation at $z=0$, constructed the simulated potential by adding up stars, dark matter, and gas mass distribution, and integrated their orbits in the potential based on their instantaneous positions and velocities. We then calculated the time-averaged $\lambda_z$ and $r$ along the orbits and took these as the particle's orbital properties. By doing this, all the particles on box orbits, which could instantaneously have a large variety of $\lambda_z$ values, will have $\lambda_z \simeq 0$ as the time-averaged values along their orbits. 

In a second approach, we adopted an approximate method \citep{zhu2018b}, whereby averaging is done in phase-space. In practice, we assumed stellar particles with similar energy, $E$, and angular momentum, $L_z$, to be on similar orbits: we hence measured the average $r$ and $\lambda_z$ across such particles from the simulation and adopted them as the orbital $r$ and $\lambda_z$ values. This method cannot return $\lambda_z =0$ for all particles on box orbits, but it does narrow their $\lambda_z$ distribution.

We integrated the stellar orbits (i.e., followed the first method) using the code of AGAMA \citep{Vasiliev2019} for four TNG50 case-study galaxies whose evolution and formation are studied in great detail (see Sect.~\ref{ss:case}), so throughout the paper their stellar orbit properties are time-averaged values. The integration takes $\sim1$ CPU hour for each galaxy. For the statistical study of the large galaxy samples from the TNG and EAGLE simulations, we adopted the second method, which is good enough for our four-component separation (see Fig.~\ref{fig:rlz_3} in the appendix).\ For the four case-study galaxies from TNG50, we can in  fact obtain galactic components via both methods, and we find consistent results. Throughout the paper, the radius of the orbit, $r$, refers to the time-averaged radius or the equivalent phase-space-averaged radius.

\begin{figure}[h!]
\centering\includegraphics[width=9cm]{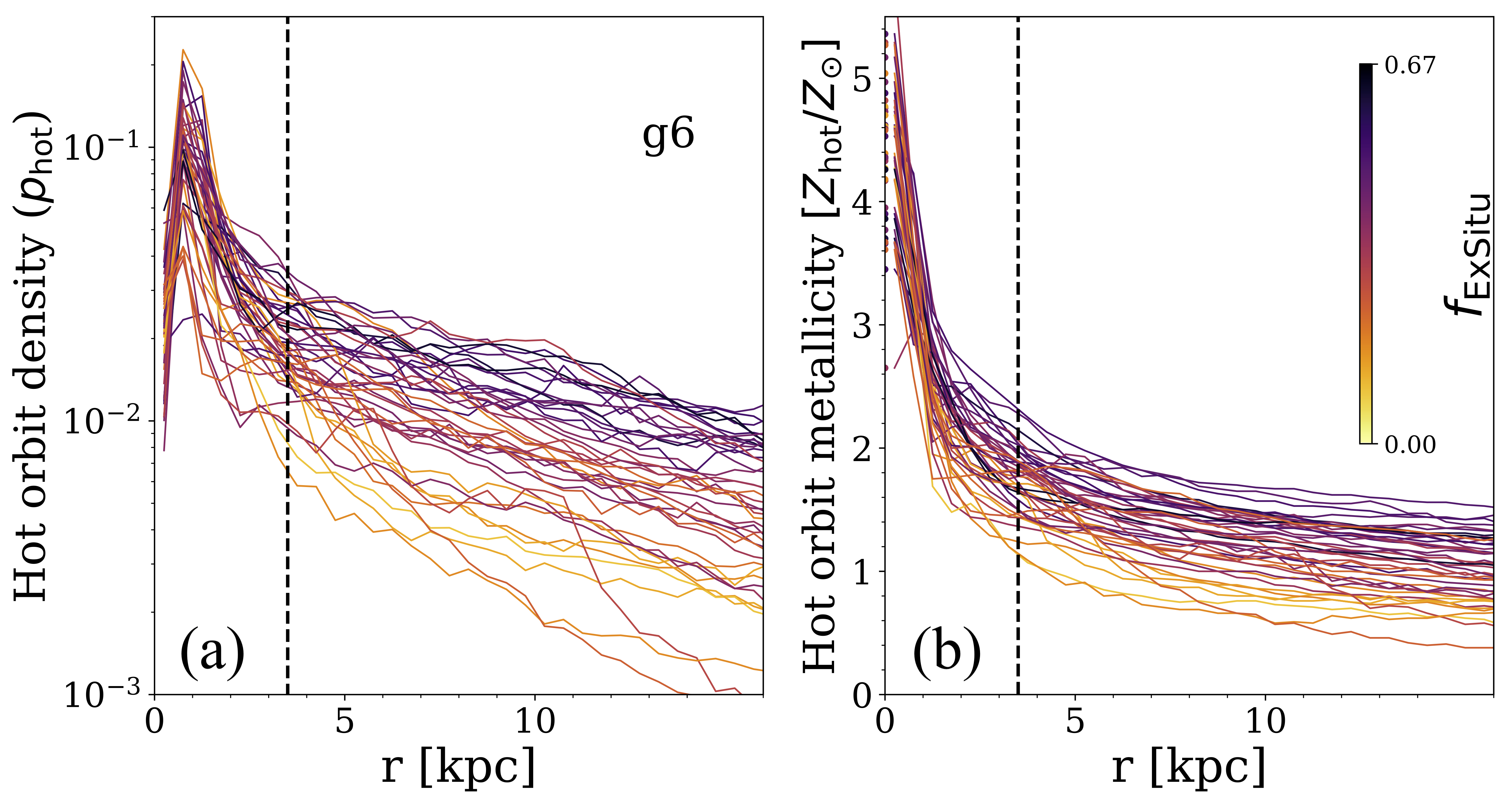}
\centering\includegraphics[width=6cm]{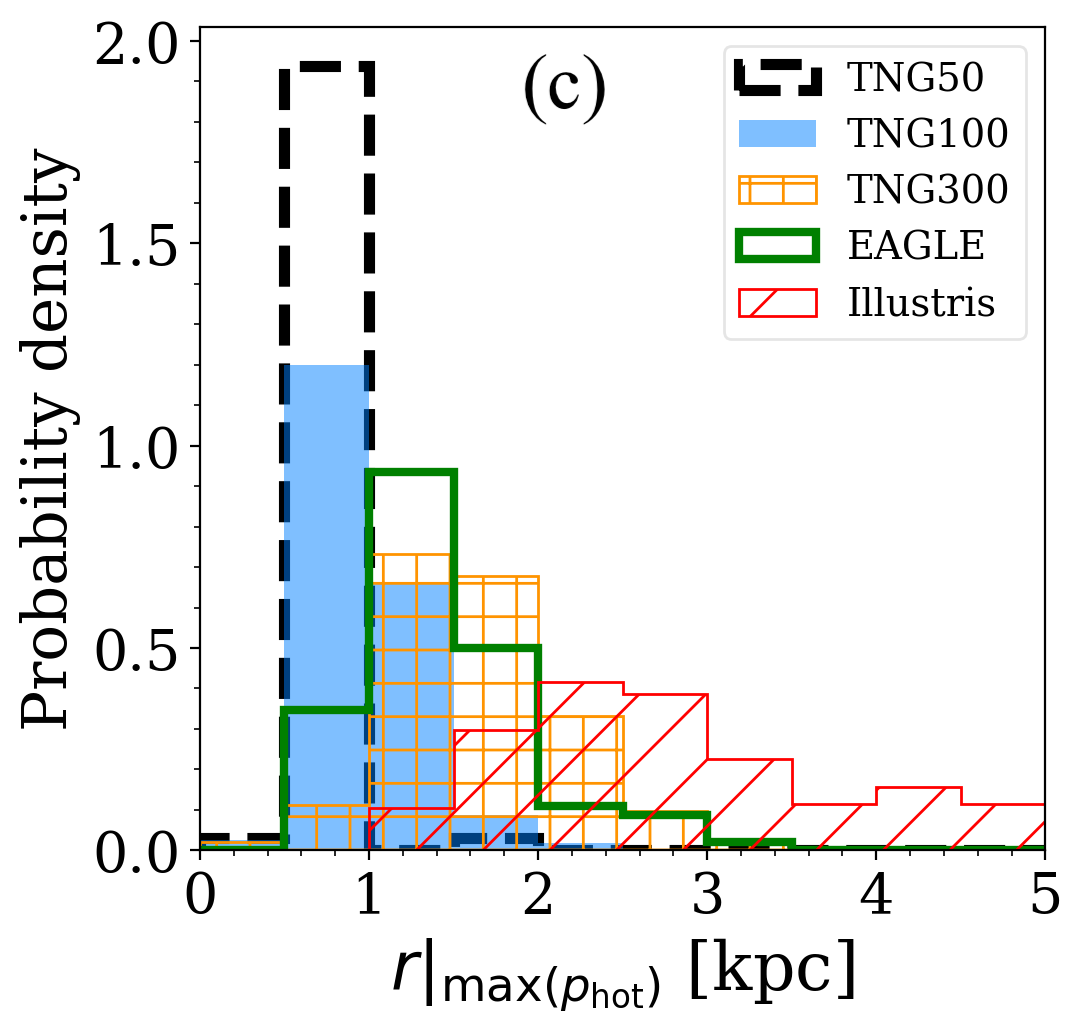}
\centering\includegraphics[width=9cm]{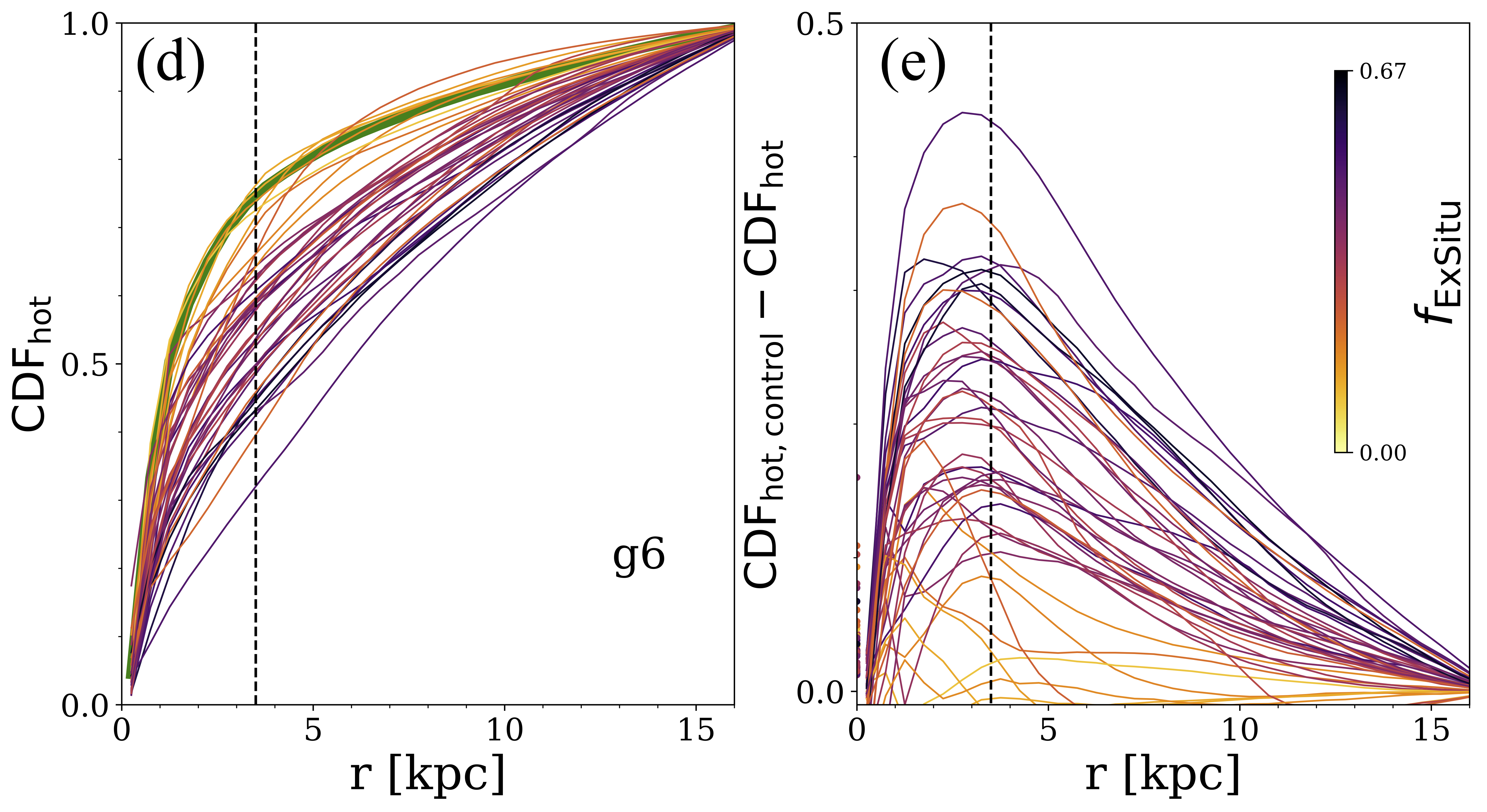}
\caption{{Properties of the stellar orbit distributions of simulated galaxies and physical justification of the definition of hot inner stellar halo.} 
{\it Top:} Probability density distribution in panel (a) and the average metallicity in panel (b), as a function of radius, $r$, obtained by summing up all the hot orbits with
$\lambda_z < 0.5$ from a selection of TNG50 galaxies at $z=0$, namely those in group g6 of Fig.~\ref{fig:MRe}. Each galaxy's curve is color coded by the galaxy's stellar ex situ fraction, as indicated by the colorbar. The vertical line indicates the radius $\rcut =
3.5$\,kpc, where both the probability density and metallicity profiles flatten. 
{\it Middle:} Distributions of the radii where the probability density distribution of hot orbits peaks, $r|_{{\rm max}(p_{\rm hot})}$, for galaxies with $M_*>10^{10.3}$\,\Msun\, in the TNG50, TNG100, TNG300, EAGLE, and Illustris simulations. Most galaxies in TNG50 have $r|_{{\rm max}(p_{\rm hot})} < 1.5$\,kpc, consistent with real galaxies from the Fornax3D survey. Galaxies in TNG100, TNG300, and EAGLE have slightly larger bulge sizes but still with $r|_{{\rm max}(p_{\rm hot})} < 3.5$ kpc.
{\it Bottom: Panel(d):} Cumulative distribution function of the hot orbits with
  $\lambda_z < 0.5$, ${\rm CDF}_{\rm hot}$ as a function of radius, $r$, for a selection of TNG50 galaxies. Symbols and selection are the same as in the top panels. The green curve represents the control model constructed by the galaxies with the lowest stellar ex situ fraction. {\it Bottom: Panel(e):} Deviation of ${\rm CDF}_{\rm hot}$ from the control model, ${\rm CDF}_{\rm hot, control}$.  The vertical line indicates the radius $\rcut = 3.5$\,kpc, which we adopt throughout as the inner boundary of the hot inner stellar halo.}
\label{fig:rcut}
\end{figure}

\subsection{Separation into four stellar components}
\label{ss:components}
\subsubsection{General separation}

In Fig.~\ref{fig:rlz_468590} we show the stellar orbit distribution, $p(r, \lambda_z)$, the stellar age distribution, $t(r, \lambda_z)$, and the stellar metallicity distribution, $Z(r, \lambda_z)$, in the phase-space of circularity, $\lambda_z$, versus radius, $r$, of a simulated galaxy from TNG50 at $z=0$ (Subfind ID 468590). We use the distributions in Fig.~\ref{fig:rlz_468590} to visualize the orbital decomposition that we use throughout the paper. 

We decomposed a galaxy into four components based on the structures in the phase-space of $\lambda_z$ versus $r$. We first notice that, in the case of ``disky'' galaxies, dynamically cold orbits with $\lambda_z > 0.8$ are usually younger and more metal rich than the other orbits. 
In the simulated galaxies, these orbits represent a rather well-defined thin stellar disk. 

For the remaining non-disk orbits, we identified the following cuts to separate hot inner stellar halos from possible bulges and ``thick'' disks. Namely, we defined the hot inner stellar halo by selecting the orbits with $r_{\rm cut}<r<r_{\rm max}$ and $\lambda_z < \lambda_{z, {\rm cut}}$; throughout this paper, $r_{\rm cut}$ is equal to 3.5 kpc, $r_{\rm max}$ is chosen based on the maximum radius covered by IFU observations in real galaxies, and $\lambda_{z, {\rm cut}} = 0.5$. 

In practice, choosing $r_{\rm cut}$ means identifying a way to separate between the bulge and the hot inner stellar halo. We discuss this choice in detail in the next subsection. 

Due to the limited observational data coverage, we can only obtain the stellar orbit distribution of real galaxies out to certain galactocentric distances. In the Fornax3D survey \citep{Sarzi2018}, the data coverage extends to $\sim 2R_e$ for most galaxies. Here we chose $\rmax = 2R_e$ for galaxies with $R_e\geq3.5\,{\rm kpc}$, whereas for galaxies with $R_e<3.5\,{\rm kpc}$ we chose $\rmax = 7\,{\rm kpc}$: these choices imply that we can focus on what people may identify or call the ``inner'' stellar halo of a galaxy, rather than the halo component that extends out to a distance of $\sim$100 kpc. 

We further distinguished between hot and warm orbits beyond the bulge radial cut by imposing a separation at $\lambda_{z, {\rm cut}} = 0.5$. The idea was to distinguish between halo stars and stars in warmer orbits that could be identified as thick disk stars and hence possibly exhibit more complex and mixed origins. In the discussion, we quantitatively show that the choice of $\lambda_{z, {\rm cut}}$ for the identification of hot orbits is not important so long as  $\lambda_{z, {\rm cut}}\lesssim 0.6-0.7$.

In summary, we separated each galaxy into four components: (i) the disk ($\lambda_z > 0.8$ and $r < r_{\rm max}$), (ii) the bulge ($\lambda_z < 0.8$ and $r < r_{\rm cut}$), the warm component ($0.5<\lambda_z < 0.8$ and $r_{\rm cut}<r < r_{\rm max}$), and the hot inner stellar halo ($\lambda_z < 0.5$ and $r_{\rm cut} <r < r_{\rm max}$).\ Here, $r_{\rm cut} = 3.5$\, kpc and $\rmax = 2R_e$ for galaxies with $R_e>3.5\,{\rm kpc,}$ and $r_{\rm max}$ is fixed at $7\,{\rm kpc}$ otherwise.

\subsubsection{Choice of $r_{\rm cut}$}
\label{sss:rcut}

In simulated and observed galaxies, we find that non-disk stellar orbits are high density and metal rich in the inner regions and become progressively lower density and more metal poor toward larger galactocentric distances. However, how to separate between the two components, the bulge and the hot inner stellar halo, and thus how to choose $r_{\rm cut}$, is not immediately obvious. In the following, we provide arguments in support for our choice, $r_{\rm cut} = 3.5$\, kpc.

In the top panels of Fig.~\ref{fig:rcut} we show the probability density distributions in panel (a) and the average metallicities in panel (b) as a function of radius, $r$, of the dynamically hot orbits (i.e., with $\lambda_z < 0.5$) in a selection of TNG50
galaxies (one galaxy, one curve). 
In particular, the diagnostics here are shown for galaxies in the group g6 (see Fig.~\ref{fig:MRe}). As can be seen, the probability densities, $p_{\rm hot}$, first decrease sharply from a peak at small radii ($<5$ kpc) and then flatten. On the other hand, the metallicities, $Z_{\rm hot}$, have their maximum at the very center of galaxies, decrease sharply at first, and then flatten. Each galaxy curve is color coded by the ex situ stellar mass fraction,
and the probability density distributions ($p_{\rm hot}$) are normalized such that the total probability of all stars within $2\,R_e$ equals unity.
At a similar galaxy mass, galaxies with a higher stellar ex situ fraction have a higher fraction of hot orbits (lower fraction of disk), whereas the central density peaks are similar and galaxies with a higher ex situ fraction exhibit a higher plateau at $r\gtrsim 3.5$ kpc in hot orbit densities and at higher metallicity values. 

We extended this characterization to all galaxies with  $10^{10.3}<M_*<10^{11.6}$\,\Msun\, in the TNG50, TNG100, TNG300, and EAGLE simulations by measuring the peaking radius, \rmaxhot, for each galaxy (i.e., the radius where $p_{\rm hot}$ reaches the maximum). Again for comparison, we also show results from Illustris galaxies. In panel (c) of Fig.~\ref{fig:rcut} we can see that most TNG50 massive galaxies have \rmaxhot $ < 1.5$\,kpc: this is consistent with measurements of real galaxies of similar mass and size from the Fornax3D survey, as we will show in the companion paper. On the other hand, galaxies in other simulations are less dense in the very inner regions, which is shown by the larger \rmaxhot\, values in Fig.~\ref{fig:rcut}: this indicates somewhat overly large bulge sizes, which can be due to a combination of a larger representation of more massive galaxies. This, in turn, is due to the larger volumes of the simulations in comparison to TNG50, lower numerical resolutions, larger force softening length, and/or different galaxy model prescriptions. In particular, a large fraction of Illustris galaxies have $\rmaxhot > 3.5$\,kpc: this is not consistent with real galaxies (see the companion paper for galaxies from the Fornax3D survey), and therefore we do not include Illustris galaxies in any further analysis. In contrast, TNG100 galaxies have \rmaxhot\, similar to TNG50 but shifted slightly to larger values, as would be expected given the lower resolution.

To quantitatively determine a physically motivated transition radius, $r_{\rm cut}$, we show in panel (d) of Fig.~\ref{fig:rcut} the cumulative distribution function of the hot orbits, ${\rm CDF}_{\rm hot}$, of the TNG50 galaxies of the top panels. We also construct a control model by using the galaxies with the lowest stellar ex situ fraction ($f_{\rm ExSitu}-\min(f_{\rm ExSitu}) < 0.05$) in the subsample. In panel (e) we show the deviation of the cumulative distribution function of each galaxy from the control model, ${\rm CDF}_{\rm hot, control} - {\rm CDF}_{\rm hot}$, again color coded by the stellar accretion fraction of the galaxy. 

The underlying idea of this comparison is as follows: we postulate that the dynamically hot orbits of a galaxy would form a pure bulge if the galaxy had no mergers, whereas we call the component that would be formed by mergers the hot inner stellar halo. The ``pure'' bulge model is unknown in detail, but we approximate it by the control model that is constructed by galaxies with the smallest ex situ fraction in the subsample. Therefore, we take the deviation from the control model to indicate a sort of perturbation caused by the mergers (i.e., the contribution to a hot inner stellar halo component).

We hence defined the transition radius between the bulge and the hot inner stellar halo as the radius ($r|_{\rm max(CDF deviation)}$) where the deviation, ${\rm CDF}_{\rm hot, control} - {\rm CDF}_{\rm hot}$, reaches its maximum. 
This is distributed closely around 3.5\,kpc for TNG50 galaxies in groups g2, g5, g6, and g9, though with larger scatter in g3, and it is slightly smaller and with larger scatter in g1, g4, g7, and g8 (see Fig.\ \ref{fig:rcdf_9} in the appendix).
Galaxies in each group have a range of sizes, $R_e$, while $r|_{\rm max(CDF deviation)}$ is distributed closely around a physical size of $\sim 3.5$ kpc. The bulges in these galaxies have similar physical sizes, consistent with the results from \cite{Du2020}; the bulge size does not scale with $R_e$, which is instead mostly set by the buildup of the outer disk or the halo.

To keep the analysis simple and easy to compare with real observations, we chose $\rcut = 3.5$\, kpc universally across all galaxies. 
The choice of $\rcut = 3.5$ is not perfect for the galaxies from the other simulations, as they exhibit larger bulges and thus somewhat larger transition radii. However, a bulge to halo separation at 3.5 kpc still applies as a well-motivated choice for TNG100, TNG300, and EAGLE galaxies with $10^{10.3}<M_*<10^{11.6}$\,\Msun, as for most of them the maximum of $p_{\rm hot}$ is reached at $\rmaxhot < 3.5$\,kpc, and thus the hot-orbit distributions $p_{\rm hot}$ are declining at 3.5 kpc. 
  
\begin{figure}
\centering
\includegraphics[width=8cm]{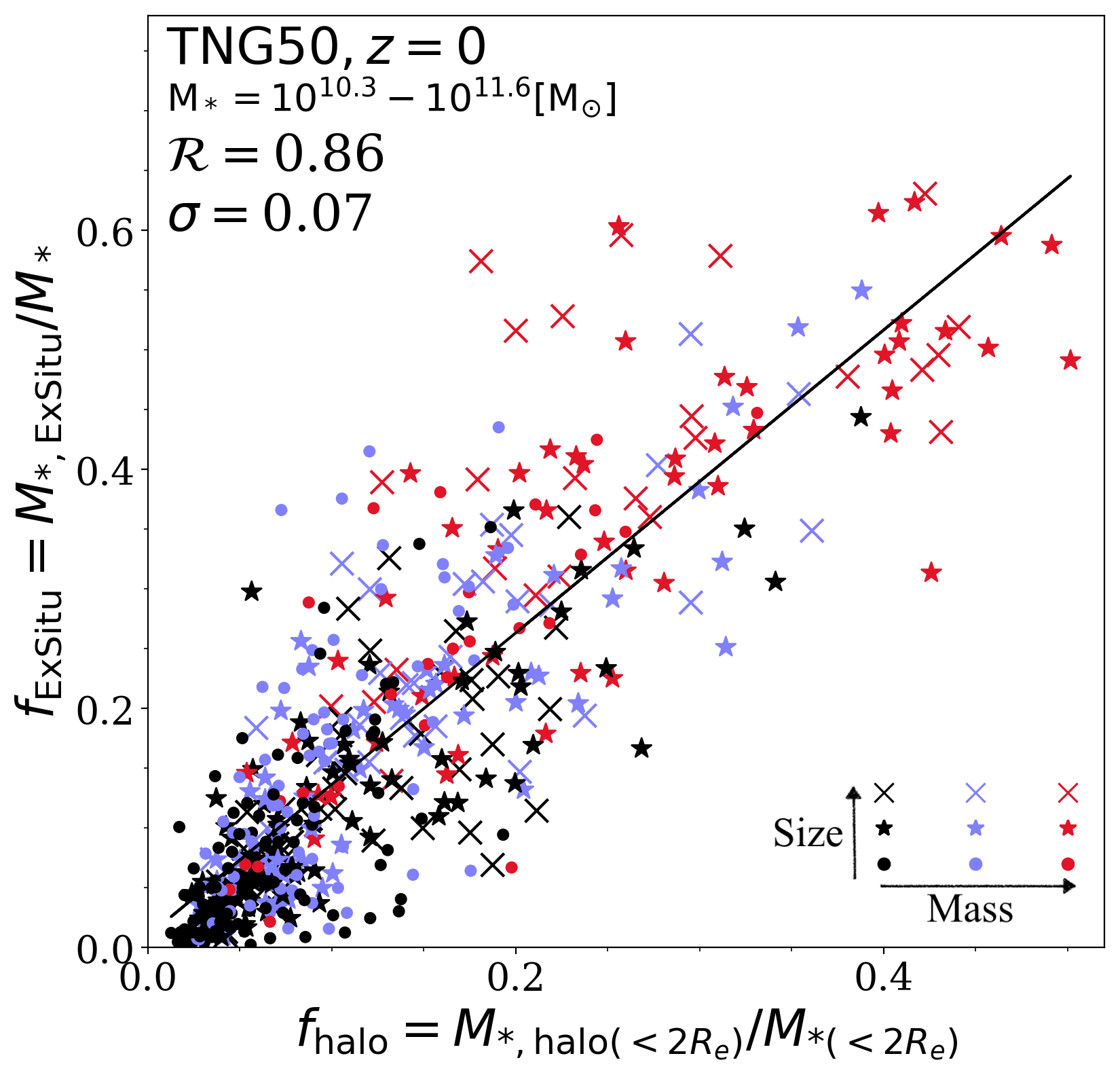}
\caption{{Connection between the hot inner stellar halo and the ex situ mass fraction according to TNG50.} We show the mass fraction of the hot inner stellar halo compared to the total stellar mass within the maximum radius of $r_{\rm max} = 2\,R_e$ ($f_{\rm halo} = M_{\rm *,halo(<2R_e)}/M_{\rm *(<2Re)}$) versus the stellar ex situ fraction ($f_{\rm ExSitu}=M_{\rm *, ExSitu}/M_*$) for galaxies with $M_*>10^{10.3}$ of TNG50 at $z=0$. Here we exclude 37 objects identified with ongoing mergers or those with counter-rotating thin disks. Galaxies with size $R_e$ below, around, and above the average are indicated by dots, asterisks, and crosses, respectively: they are color coded based on stellar mass, with increasing stellar mass from black, to blue, to red. Galaxies with larger stellar masses and sizes exhibit systematically larger ex situ fractions as well as larger mass fractions in the hot inner stellar halo. The Pearson correlation coefficient $\mathcal{R}(f_{\rm halo}, f_{\rm ExSitu}) = 0.86$ of the whole sample is annotated. The black line is a linear fit to the data points, while the $1\sigma$ scatter of the data points against the linear fit is $\sigma = 0.07$. } 
\label{fig:fhalo_fexsitu}
\end{figure}

\begin{figure*}
\centering\includegraphics[width=18cm]{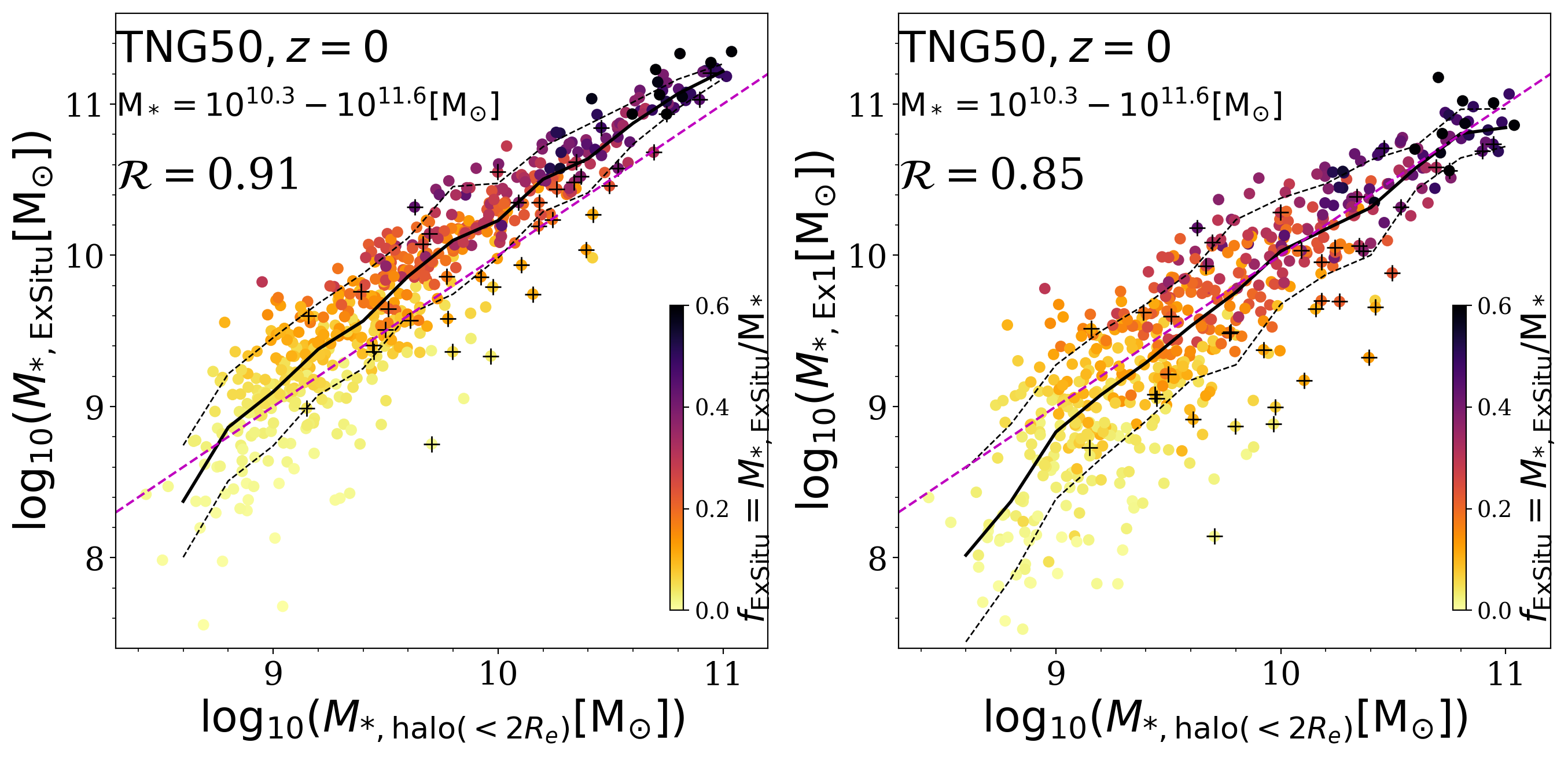}
\caption{
{Correlations between the mass of the hot inner stellar halo,
\Mshalo, and the total accreted stellar masses (left) and the stellar mass accreted from the most massive merger (right), according to TNG50}. 
In the left panel, $M_{*,\mathrm{ExSitu}}$ includes all the ex situ stellar mass ever accreted from all past mergers and from the stripping of ongoing mergers, satellites, and flybys. In the right panel, $M_{*,\mathrm{ExSitu}}$ is the stellar mass accreted from the most massive merger.
Here all galaxies with $M_*>10^{10.3}$ from TNG50 at $z=0$ are included, color coded by their stellar ex situ fraction. The thick black curve is the running median, and the thin dashed curves represent the $\pm1\sigma$ scatter. Black plus symbols mark galaxies with ongoing merging or those with counter-rotating disks. The Pearson correlation coefficients, $\mathcal{R}$, of the correlations are annotated in each panel. The dashed magenta line refers to 1:1. Some satellite mergers will deposit their stellar mass across a wide radius, with only a fraction of it within $2R_e$, and so can lie above that line.
On the other hand, the x axis shows a combination of in situ and ex situ stars such that the total mass can greatly exceed the total mass of the satellites in some galaxies with low ex situ fractions, causing them to lie below the line. }
\label{fig:Macc3_tng50_1}
\end{figure*}

\section{Information content of the hot inner stellar halo}
\label{s:Results}

We took all the galaxies with $10^{10.3}<M_* <10^{11.6}$\,\Msun\, in the TNG50, TNG100, TNG300, and EAGLE simulations at $z=0$. For each of them, we measured the mass of their hot inner stellar halo, \Mshalo, by adopting the orbital decomposition described in Sect.~\ref{ss:components}: we did so irrespective of their global morphology, kinematics, star formation properties, and environments. 

We also kept track of the merger and stellar accretion histories of each simulated galaxy and summarized them by means of statistics, such as the total ex situ stellar mass.\ This is the total amount of stellar mass accreted via mergers and via the stripping of satellites (see Sect.~\ref{sec:TNG}) and the galaxy stellar mass of the most massive progenitor ever merged with.

In the following we show that the mass of the hot inner stellar halo alone is sufficient to infer selected summary statistics of the merger and assembly history of individual galaxies. 

\subsection{Mass fraction of the hot inner stellar halo}
\label{SS:fhalo}
In Fig.~\ref{fig:fhalo_fexsitu} we show how the mass fraction of the hot inner stellar halo is connected to the galaxy's stellar ex situ fraction, $f_{\rm ExSitu}$.
We define the mass fraction of the hot inner stellar halo compared to the total stellar mass within the maximum radius of $r_{\rm max} = 2R_e$ as $f_{\rm halo} = M_{\rm *,halo(<2R_e)}/M_{\rm *(<2Re)}$.

By checking the stellar images of all TNG50 galaxies in the mass range shown in the figure (with $M_* = 10^{10.3-11.6}$\,\Msun\,), we identified 37 that are undergoing a merger or exhibit a counter-rotating thin disk at the time of inspection ($z=0$). Galaxies with ongoing mergers are not in dynamical equilibrium, making a meaningful definition of the hot inner stellar halo impossible. On the other hand, counter-rotating thin disks are usually formed in situ from  misaligned gaseous disks, regardless of mergers. These 37 galaxies significantly deviate from the correlation and are excluded from the figure. 

Mass fraction of the hot inner stellar halo, $f_{\rm halo}$, is strongly correlated with the galaxy's stellar ex situ fraction, $f_{\rm ExSitu}$.
We get a Pearson correlation coefficient of $\mathcal{R}(f_{\rm halo}, f_{\rm ExSitu}) = 0.86$ with the data points in the figure, and it decreases to 0.82 when the 37 objects are included.
The black line is a linear fit to the data points, and we calculate that the $1\sigma$ scatter of the data points against the linear fit is $\sigma = 0.07$.  

In Fig.~\ref{fig:fhalo_fexsitu} we show that galaxies with larger stellar mass, $M_*$,  and size, $R_e$, have systematically larger $f_{\rm ExSitu}$ and $f_{\rm halo}$. Thus, the variation in $f_{\rm ExSitu}$ across the mass-size plane (as shown in Fig.~\ref{fig:MRe}) is encoded in the variation in $f_{\rm halo}$. Moreover, galaxies with similar stellar mass and size still span a wide range in $f_{\rm ExSitu}$ and $f_{\rm halo}$. The relation we show here reflects the physical connection between the increase in the stellar ex situ fraction and the growth of the hot inner stellar halo, regardless of the galaxy stellar mass and size.

\subsection{Correlations between the hot inner stellar halo mass and the ancient accreted stellar mass}
\label{ss:c1}

Figure~\ref{fig:Macc3_tng50_1} gives the main results of this paper. There, we show the total ex situ stellar mass, $M_{*, {\rm ExSitu}}$ (left), and stellar mass accreted from the most-massive galaxy ever merged by TNG50 galaxies, $M_{*, {\rm Ex1}}$ (right), as a function of their hot inner stellar halo mass, \Mshalo. Galaxies are color coded by their total stellar ex situ fraction.\footnote{Here, the total ex situ stellar mass includes stars accreted from all galaxies that have ever merged and from all satellites still orbiting or flying by at the time of inspection. By ``most massive galaxy ever merged'' we mean the most massive galaxy that has in-fallen into the galaxy under scrutiny and that, by the time of inspection at $z=0$, no longer exists as an individual object, having been destroyed by the gravitational interaction with the galaxy under scrutiny. Throughout the paper the most massive merger's mass, $M_{*, {\rm Ex1}}$, is defined and measured as all the stellar mass accreted from the most massive secondary progenitor of a galaxy that merged with the host. It could be larger than the galaxy stellar mass of the satellite at any time prior to the merger because the merging object could be continuously losing as well as forming new stars during the merger.}


\begin{figure*}
\centering\includegraphics[width=18cm]{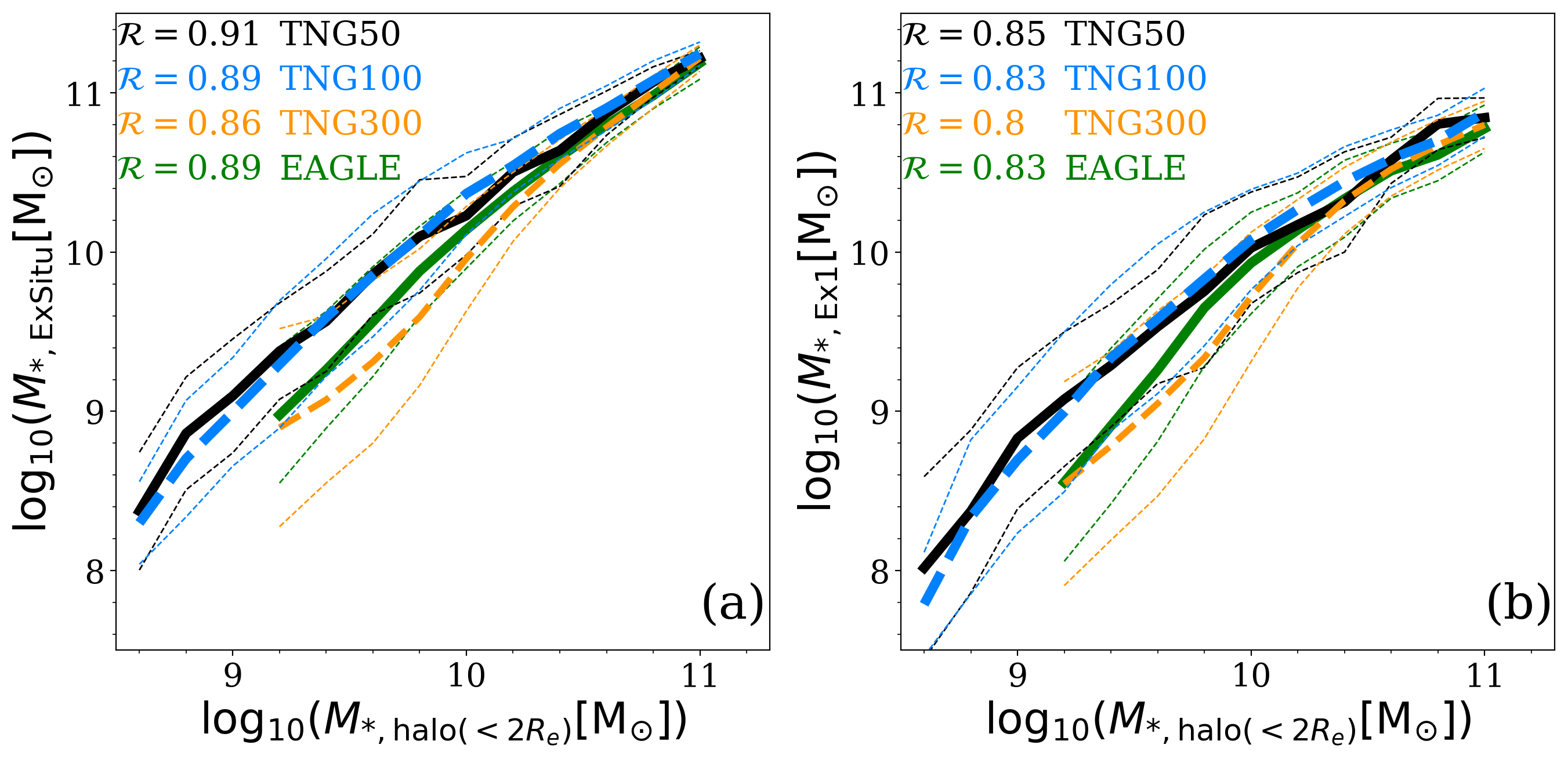}
\caption{{Information content of the hot inner stellar halo across galaxy formation simulations.}
We show the correlations of \Mshalo\, with the total ex situ stellar mass, $M_{*,\mathrm{ExSitu}}$ (left), and with the stellar mass accreted from the most massive merger, $M_{*,\mathrm{Ex1}}$ (right), for different galaxy simulations. 
The thick black, blue, yellow, and green lines are the median curves from TNG50, TNG100, TNG300, and EAGLE galaxies at $z=0$, respectively, in the $M_* = 10^{10.3-11.6}$\,\Msun\, range; the thin curves are the $\pm1\sigma$ scatters. The Pearson correlation coefficients, $R$, from the four sets of simulations are labeled in the corresponding colors.
}
\label{fig:Macc_4sims}
\end{figure*}

It is manifest from Fig.~\ref{fig:Macc3_tng50_1} that the hot inner stellar halo mass is tightly correlated with the total ex situ stellar mass, with Pearson correlation coefficient $\mathcal{R}(\Mshalo, M_{*, {\rm ExSitu}}) = 0.91$, as well as with the stellar ex situ fraction. 
The correlation with the stellar mass accreted from the most massive progenitor a galaxy has ever merged with is slightly weaker, with $\mathcal{R}(\Mshalo, M_{*, {\rm Ex1}}) = 0.85$.

The $1\sigma$ scatter of the correlations against the running medium is $\sim0.5$ dex at $\Mshalo < 10^{10}$\,\Msun. It becomes as small as $\sim 0.1$ dex for $M_{*,\mathrm{ExSitu}}$ and $\sim 0.15$ dex for $M_{*,\mathrm{Ex1}}$ at the high-mass end explored here, whereby $\Mshalo \sim 10^{11}$\,\Msun\, (i.e., a total galaxy stellar mass of $10^{11.6}$\,\Msun). 

The 37 galaxies that are undergoing a merger or exhibit a counter-rotating disk at the time of inspection ($z=0$) are included in Fig.~\ref{fig:Macc3_tng50_1}: they are typically outliers, most of them lying below the median trend.
The  Pearson correlation coefficients $\mathcal{R}(\Mshalo, M_{*, {\rm ExSitu}})$ and $\mathcal{R}(\Mshalo, M_{*, {\rm Ex1}})$ of the whole TNG50 sample increase to 0.92 and 0.86, respectively, when the 37 galaxies are excluded. Apart from this, most outliers lying below the correlations are galaxies with a stellar ex situ fraction $f_{\rm ExSitu} < 0.1$. 

It is important to highlight that the correlation between the hot inner stellar halo mass and the total ex situ stellar mass (or the mass of the most massive merger) is stronger than the one between the latter and the total stellar mass of a galaxy: this is the case even though all quantities of Fig.~\ref{fig:Macc3_tng50_1} correlate with a galaxy's mass. This can be learned from Fig.~\ref{fig:fhalo_fexsitu}, whereby it is manifest that the mass fraction of the hot inner stellar halo is highly correlated with the stellar ex situ fraction. We also quantify this in Appendix C: for comparison, for the same TNG50 galaxies, the Pearson correlation coefficient between a galaxy stellar mass and $M_{*, {\rm Ex1}}$ reads 0.75 and the $1\sigma$ scatter varies from 0.6 to 0.3 from the low-mass to the high-mass end (see Fig.~\ref{fig:mstar_4sims}).

In Fig.~\ref{fig:Macc_4sims} we demonstrate that the existence of tight correlations is a prediction not exclusive to TNG50: we see that there are similar correlations for TNG50 (repeated from Fig.~\ref{fig:Macc3_tng50_1}), TNG100, TNG300, and EAGLE galaxies. In the left panel, we show the relationships between the hot inner stellar halo mass of a galaxy and the total ex situ stellar mass, and on the right we give the relationship between the former and the accreted stellar mass from its most massive merger. 

From these comparisons we can tell that results from models with different galaxy-formation physics and resolution agree to a good degree, particularly toward the high-mass end. 
In particular, the correlations from TNG100 galaxies match those from TNG50 perfectly, with very similar median and scatter throughout the mass range explored here. With lower resolution than TNG50, TNG100 galaxies have less-dense centers and larger bulge sizes than TNG50 galaxies, as discussed with Fig.~\ref{fig:rcut}. Here we use the same $r_{\rm cut} = 3.5$\,kpc for all galaxies to separate the hot inner stellar halo from the bulge, even though the actual transition radius of TNG100 galaxies may be somewhat larger. The fact that the correlation also holds for TNG100 galaxies is an indication that the precise choice for the inner boundary of the hot inner stellar halo is not crucial so long as the structural properties of the simulated galaxies are reasonably realistic. 
 
At even lower resolution and with larger softening length, TNG300 galaxies have even larger bulge sizes than TNG100 galaxies (again, Fig.~\ref{fig:rcut}, middle panel). Our fixed choice of $r_{\rm cut} = 3.5$\,kpc causes an overestimate of the hot inner stellar halo mass, especially in galaxies with low ex situ fractions. The trends from TNG300 galaxies are consistent with those from TNG50 and TNG100 galaxies at $\Mshalo \gtrsim 10^{10.2}$\,\Msun\ ( i.e., galaxy stellar mass $\gtrsim 10^{11}$\,\Msun\,) but are steeper at lower masses. 
 
\begin{figure*}
\centering\includegraphics[width=18cm]{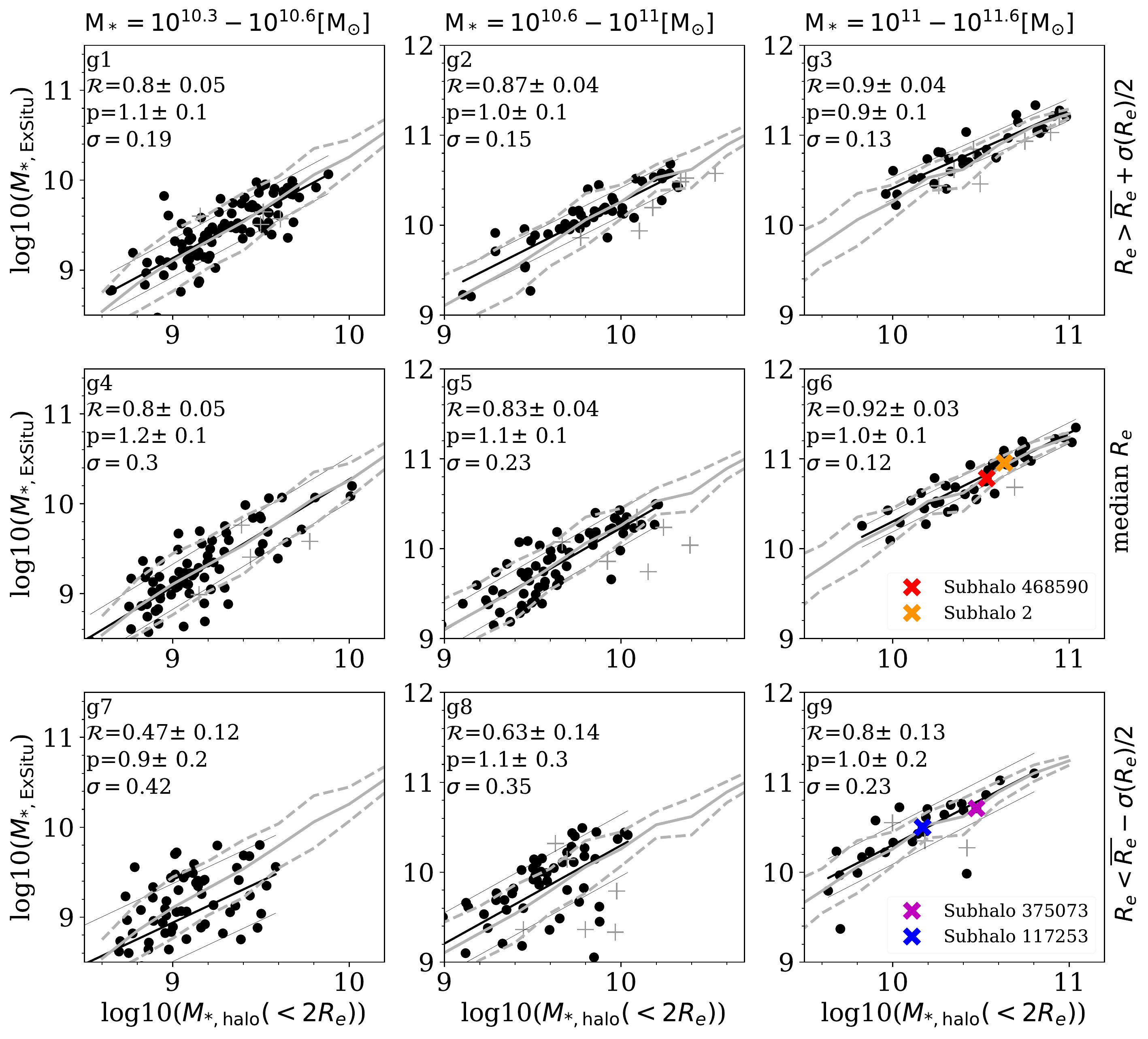}
\caption{{Correlation between \Mshalo\, and the total ex situ stellar mass, $M_{*,\mathrm{ExSitu}}$, as a function of galaxy stellar mass and stellar size.} Here we show the relationships for TNG50 galaxies divided into nine groups by mass and size, as shown in
Fig.~\ref{fig:MRe}, one grouping per panel, with galaxy stellar mass increasing from left to right and galaxy stellar size increasing from bottom to top. In the middle row, galaxies of median size for their mass are depicted. The black dots denote TNG50 galaxies without ongoing mergers, whereas the gray plus symbols indicate ongoing mergers. The thick black line is a linear fit to the black dots, and the thin black lines represent the $\pm1\sigma$ scatter. The median and $\pm1\sigma$ scatter of the whole sample as shown in Fig.~\ref{fig:Macc3_tng50_1} are reproduced as gray curves for reference. The medium and $\pm1\sigma$ uncertainty of the Pearson correlation coefficient, $\mathcal{R}(\Mshalo, M_{*,\mathrm{ExSitu}})$, slope of the linear fit, $p$, and the typical $1\sigma$ scatter of the data points against the linear fit are labeled for each group. The four galaxies that are discussed as case studies in Sect. 5 are marked with colored symbols (in panels g6 and g9). 
}
\label{fig:Macc3_9}
\end{figure*}

\begin{figure*}
\centering\includegraphics[width=18cm]{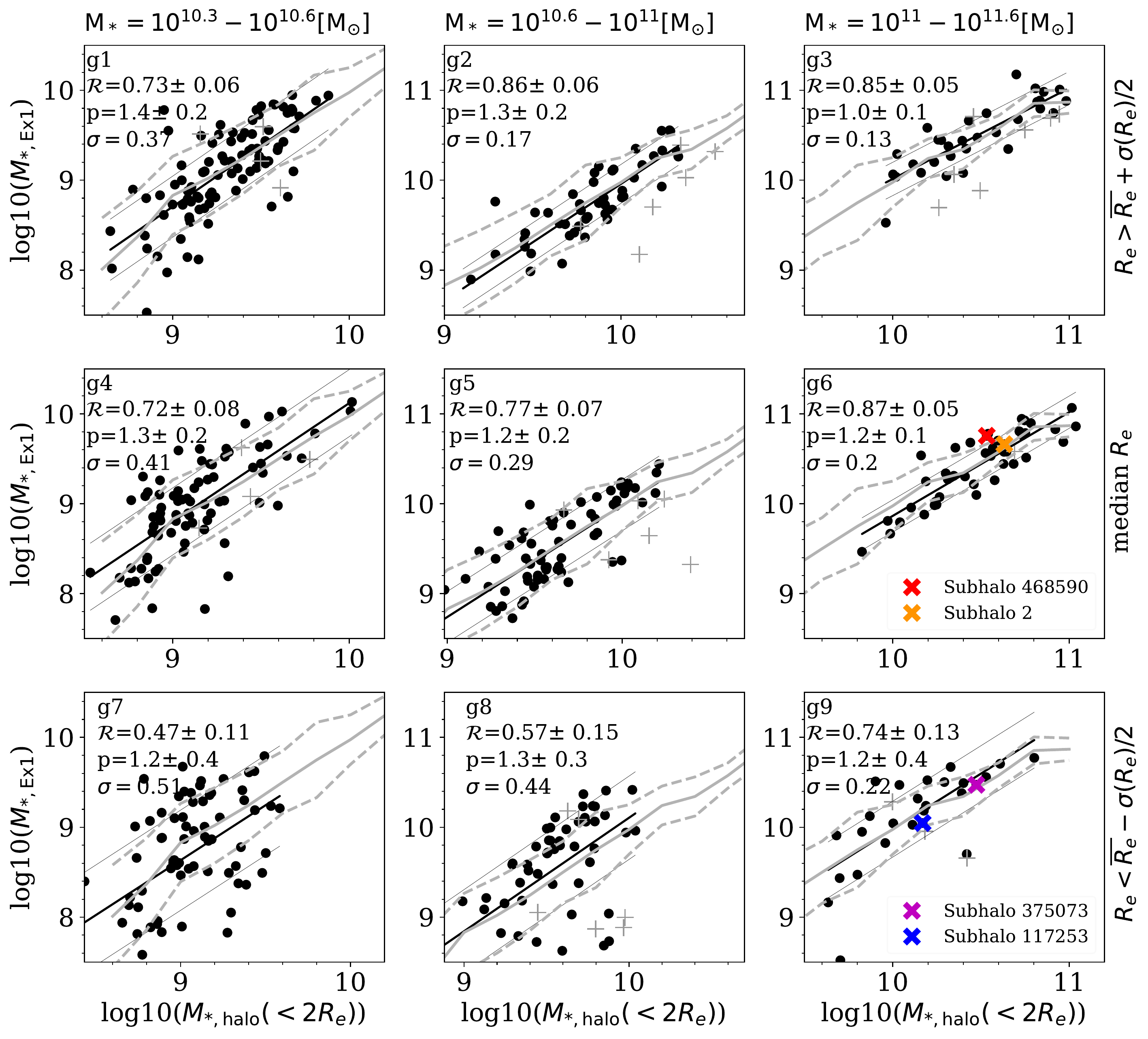}
\caption{ {Correlation between the hot inner stellar halo mass, \Mshalo, and the stellar mass accreted from the most massive merger, $M_{*,\mathrm{Ex1}}$, as a function of galaxy mass and size.} Annotations are as in Fig.~\ref{fig:Macc3_9}, and results are from TNG50.}
\label{fig:Macc_9}
\end{figure*}

EAGLE galaxies are the result of different feedback physics but have numerical resolution comparable to TNG100 galaxies and exhibit bulge sizes between those of TNG100 and TNG300 galaxies. The EAGLE correlations between the hot inner stellar halo mass and the ex situ stellar mass (total or from the most massive merger) are perfectly consistent with those from the TNG50 and TNG100 simulations at $\Mshalo \gtrsim 10^{9.8}$\,\Msun\ (i.e., galaxy stellar mass $\gtrsim 10^{10.6}$\,\Msun\,) but become steeper at $\Mshalo \lesssim 10^{9.8}$\,\Msun. Yet, the EAGLE predictions are within the $1\sigma$ scatters of the TNG50 and TNG100 predictions.

In fact, and importantly, all models predict tight correlations between the mass of the hot inner stellar halo and the total ex situ mass or the stellar mass of the most massive progenitor, with Pearson  correlation coefficients larger than 0.86 and 0.8, respectively, across all simulation runs. The trends predicted by TNG50, TNG100, and EAGLE are consistent within the $\pm1\sigma$ scatter across the explored mass range.

\subsection{Correlations for galaxies of different mass and size}
\label{ss:c9}

In Fig.~\ref{fig:MRe} we show that the ex situ stellar mass fraction
of galaxies increases with increasing stellar mass and stellar size. In fact, even when the galaxy population is divided into subsamples at fixed galaxy stellar mass and varying size (see the nine groupings of Fig.~\ref{fig:MRe}), non-negligible variations in ex situ fraction, and hence merger histories, are manifest. 
Moreover, Fig.~\ref{fig:Macc3_tng50_1} shows that the relationship between the hot inner stellar halo mass and the total ex situ stellar mass is tighter the larger the galaxy mass. In the following, we quantify how the correlations uncovered in this paper change as a function of galaxy mass and stellar size. 

In Fig.~\ref{fig:Macc3_9} we show the correlations of the hot inner stellar halo, \Mshalo, versus $M_{*, {\rm ExSitu}}$ for the subsamples of TNG50 galaxies identified in Fig.~\ref{fig:MRe}: from less to more massive galaxies from left to right, from larger to smaller galaxies from top to bottom, with the galaxies in the middle row exhibiting median stellar sizes given their stellar mass. We performed the linear fitting and calculation of the Pearson correlation coefficient many times by bootstrapping. The medium and $\pm1\sigma$ uncertainty of the Pearson correlation coefficient, $\mathcal{R}(\Mshalo,M_{*,\mathrm{ExSitu}})$, and slope, $p$, of the linear fitting are labeled in each panel.
The typical $1\sigma$ scatter of data points against the linear fit are also labeled. 

Figure~\ref{fig:Macc3_9} demonstrates that the information content of the hot inner stellar halo is maximal for the most massive and largest galaxies (i.e., the correlations are strongest in the subsamples with the most massive galaxies and/or those with a large size). The $1\sigma$ scatter is $\sim 0.12$ dex in g2, g3, and g6: these values are similar to the scatter inferred from the whole galaxy sample at the high-mass end. On the other hand, the correlations are weaker in galaxies with lower masses and smaller sizes. The $1\sigma$ scatter increases to $\sim 0.25$ dex in g1 and g5, and to $0.3-0.4$ dex in g4, g7, g8, and g9 (i.e., mostly for galaxies below the average mass-size relation).  

Figure~\ref{fig:Macc_9} shows the analog correlation of the hot inner stellar halo mass, \Mshalo, versus the stellar mass accreted from the most massive merger, $M_{*, {\rm Ex1}}$, for each subsample of TNG50 galaxies. The correlations in most groups are slightly weaker when only the most massive merger is considered (Fig.~\ref{fig:Macc_9} versus Fig. \ref{fig:Macc3_9}). The $1\sigma$ scatter is $\sim 0.2$ dex in g2, g3, and g6 and increases to $\sim 0.4-0.5$ dex in groups of galaxies with lower masses and smaller sizes. 

Importantly, Figs.~\ref{fig:Macc3_9} and \ref{fig:Macc_9} show that the average relationships between the mass of the hot inner stellar halo and the two studied summary statistics of the merger history of galaxies hold practically unchanged across the investigated galaxy population, barring the $10^{10.3-10.6}$\,\Msun\, galaxies with below-average stellar sizes. The slopes of the linear fits of the nine groups are consistent with one another within $1\sigma$ uncertainty.
On the other hand, the constraining power of the hot inner stellar halo mass is highest for more massive and more extended galaxies. 

There are still positive, albeit very weak, correlations between the total stellar mass, \Mstar\,, on the one side and $M_{*, {\rm ExSitu}}$ or $M_{*, {\rm Ex1}}$ on the other, including in narrow bins of galaxy stellar mass and size. The correlations in \Mshalo\, versus $M_{*, {\rm ExSitu}}$ (or $M_{*, {\rm Ex1}}$) are much stronger.

\section{Discussion}

We want to determine the physical origin of the relationships quantified and uncovered in Figs.~\ref{fig:Macc3_tng50_1}-\ref{fig:Macc_9} as well as the origin of the lingering galaxy-to-galaxy variations, which are also at fixed galaxy mass and size. To address these questions, we delve into the simulation data, follow the formation of the hot inner stellar halo of selected TNG50 galaxies, and check whether other properties can be used to provide additional information into the past assembly history of a galaxy. Before that, however, we comment on how the correlations presented in this paper depend on the adopted operational definition of a hot inner stellar halo. 

\begin{figure}
\centering\includegraphics[width=8cm]{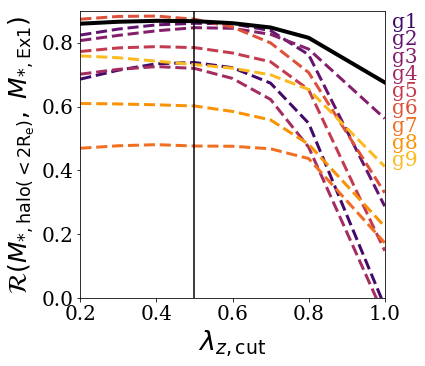}
\caption{{Dependence of the correlation between the masses of the most massive merger and of the hot inner stellar halo of a galaxy on the operational definition of the latter.} We show the Pearson correlation coefficient between $M_{*,\mathrm{Ex1}}$ and \Mshalo\, as a function of $\lambda_{z,\mathrm{cut}}$, the maximum level of orbital ``coldness'' adopted to define inner-halo stars. Namely, in the plot we adopt different definitions of hot inner stellar halo by imposing $\lambda_z < \lambda_{z,\mathrm{cut}}$.
Dashed curves in different colors indicate galaxies in subsamples g1-g9, and the solid black curve indicates galaxies of the whole sample.}
\label{fig:lcut}
\end{figure}

\subsection{Dependence on the definition of hot inner stellar halo} 
\label{SS:lcut}
In Sect.~\ref{ss:components} we chose a definition of hot inner stellar halo such that the circularity of its stars (or stellar orbits) is smaller than a given value, $\lambda_{z,\mathrm{cut}}=0.5$. Namely, in our operational definition of (inner) stellar halo, we select only for hot orbits and exclude possibly ``warm'' and ``cold'' ones -- as a reminder, the disk component is composed of cold orbits with circularity $\lambda_z > 0.8$. Here we show how the correlations discovered above depend on such a choice. 

In Fig.~\ref{fig:lcut} we show the Pearson correlation coefficient, $\mathcal{R}(\Mshalo, M_{*,\mathrm{Ex1}}$), as a function of $\lambda_{z,\mathrm{cut}}$ for TNG50 galaxies, with the latter varying between about 0.2 and 1. We study results for both the whole TNG50 sample (see the selection in Sect.~\ref{s:selection}) and the TNG50 galaxies binned by galaxy stellar mass and size (see groupings from Fig.~\ref{fig:MRe}). 
The correlation is dramatically stronger with $\lambda_{z,\mathrm{cut}} \lesssim 0.8$ for most galaxy subsamples. So long as $\lambda_{z,\mathrm{cut}} \lesssim 0.6-0.7$, the correlations are similarly informative to within $10-20$ per cent, irrespective of galaxy properties. 
The operational definition of the hot inner stellar halo adopted throughout this analysis (with $\lambda_{z,\mathrm{cut}}=0.5$) is a favorable one for the scientific questions of interest but is not fine-tuned. 

\subsection{Formation of the hot inner stellar halo}
\label{ss:case}

\begin{figure*}
\centering\includegraphics[width=14cm]{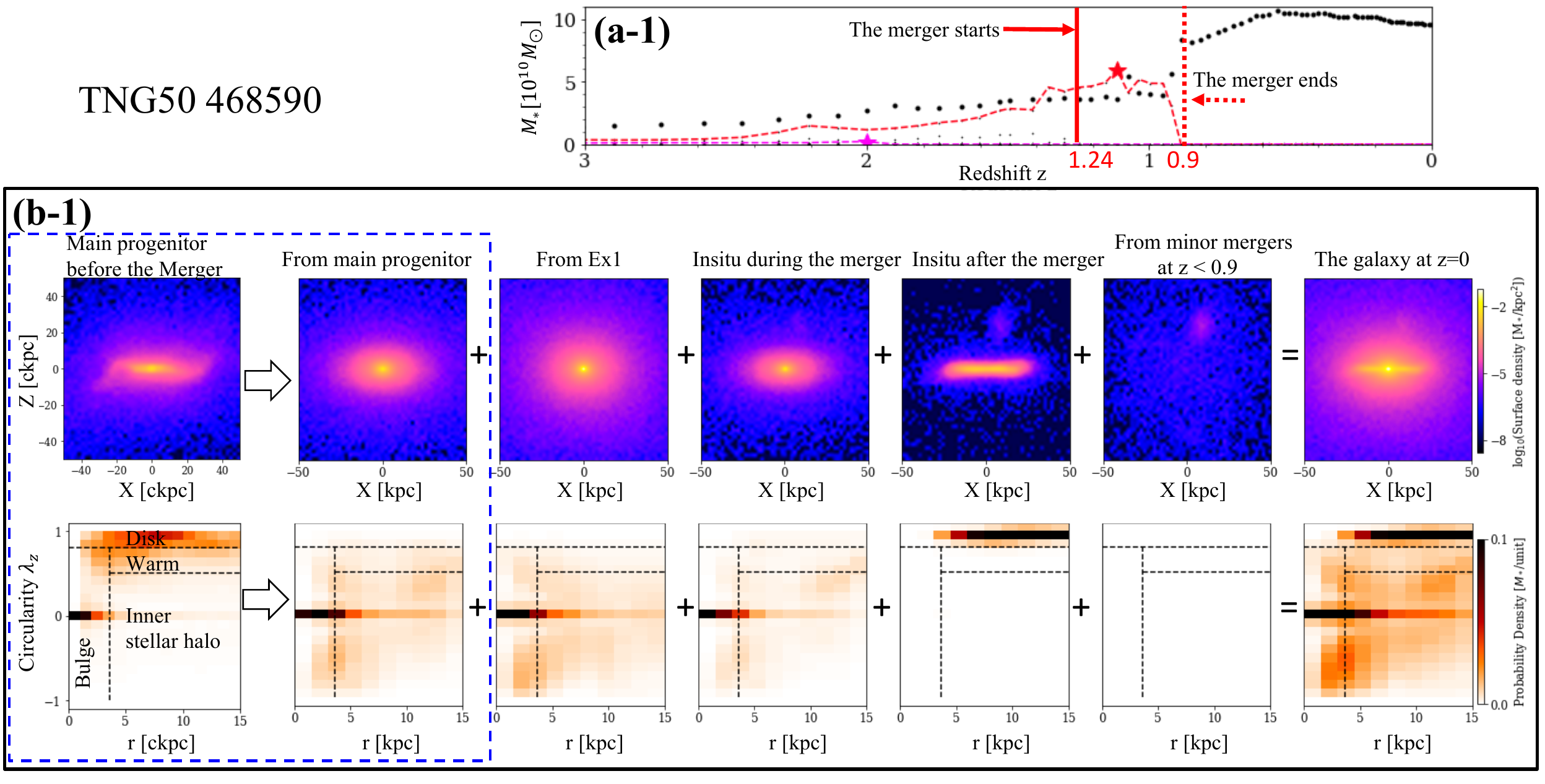}
\centering\includegraphics[width=14cm]{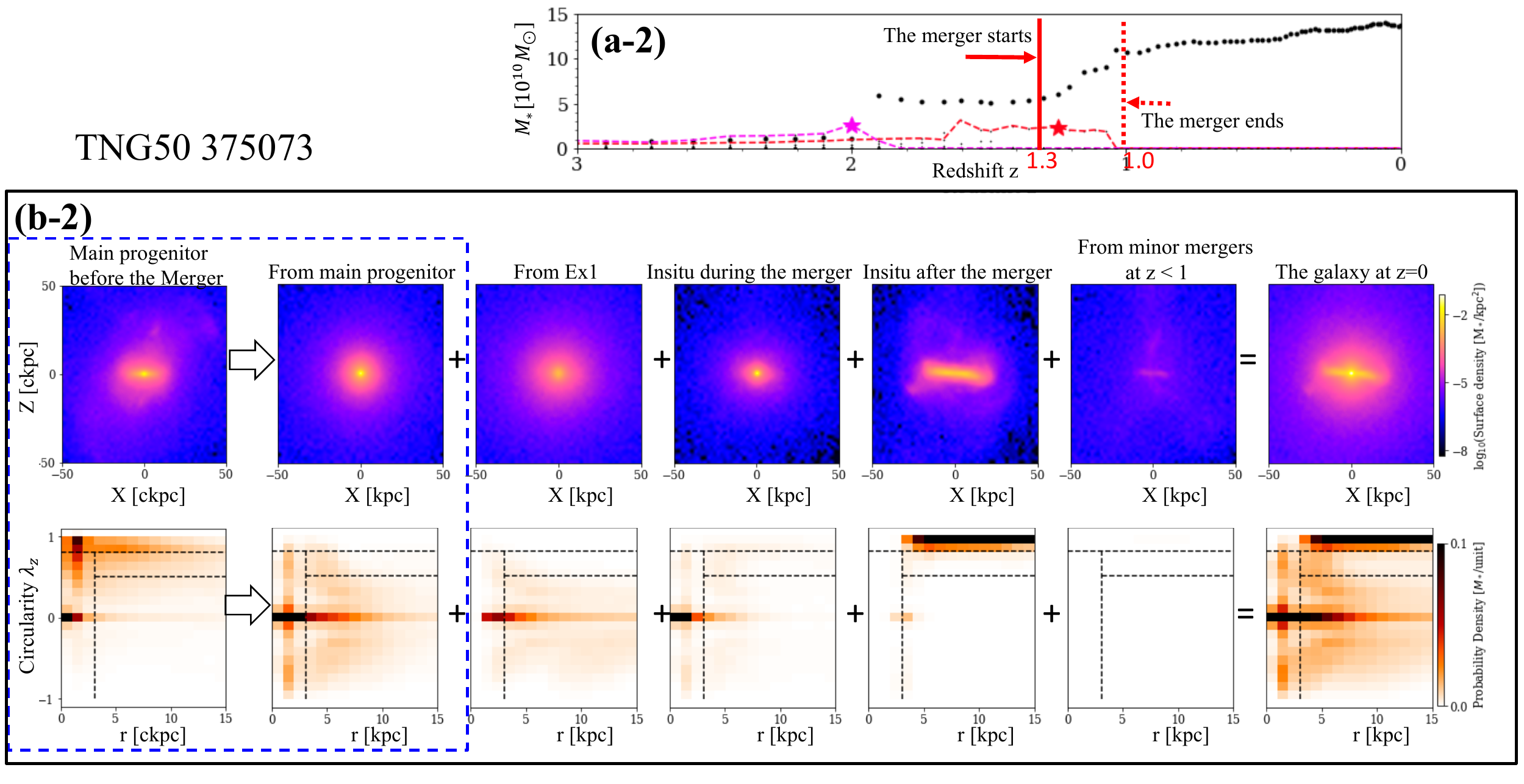}
\caption{
{Formation of the hot inner stellar halo in TNG50\,468590 (a-1 and b-1; TNG50\,468590 is the same galaxy as illustrated in Fig.~\ref{fig:rlz_468590}) and TNG50\,375073 (a-2 and b-2). In}
{panel  (a-1)} we show the stellar mass assembly history of TNG50\,468590. The black dots indicate the stellar mass evolution of the main progenitor as a function of redshift, $z$. The dashed colored lines denote the mass assembly of the secondary galaxies that merged into the main progenitors directly and at one point reached a maximum stellar mass larger than $10^{9}\,M_{\odot}$; the corresponding colored star symbols indicate the times when the maximum mass was reached. TNG50\,468590 experienced a massive merger at $z\sim 1$ with a stellar mass ratio of $\sim 1:1$. The solid and dashed red vertical lines indicate the beginning and ending of this merger. 
{In the top row of panel (b-1)}, we show the morphology of the main progenitor before the major merger in the leftmost panel and the morphology at redshift $z=0$ in the rightmost panel; in between we have the contribution to the $z=0$ morphology from different origins with (from left to right) the stars from the main progenitor, accreted from the most massive merger, formed during the major merger, formed after the major merger ends, and accreted from subsequent minor mergers. We use co-moving distances in units of kpc (labeled as ckpc) for galaxies at $z>0$. 
The surface density is in units of stellar mass per unit area kpc$^2$.
The stellar mass is normalized such that the total stellar mass of the galaxy at $z=0$ equals unity within the figure coverage.
The bottom row shows the orbital distribution of the corresponding stars in the phase-space of circularity, $\lambda_z$, versus radius, $r$, as derived from the simulation particle information. The probability density is in units of stellar mass per unit area in the phase-space.  
The dashed lines indicate our orbital-based division into four components as adopted for NGC\,1380: disk, warm component, bulge, and hot inner stellar halo. 
It has a hot inner stellar halo mass of $3.4\times10^{10}\,M_{\odot}$, and $90\%$ of it was produced by the most massive merger event. The contribution from subsequent minor mergers is negligible.
{Panels (a-2) and (b-2) } are similar but for TNG50\,375073. TNG50\,375073 had two massive mergers, one at $z\sim2$ and one at $z\sim1$. In panel (b-2), we trace the structure formation during the most massive merger, which happened at $z\sim1$. It has a hot inner stellar halo mass of  $3.0\times10^{10}\,M_{\odot}$, and $84\%$ of it was produced by the most massive merger event. 
}
\label{fig:formation1}
\end{figure*}

\begin{figure*}
\centering\includegraphics[width=14cm]{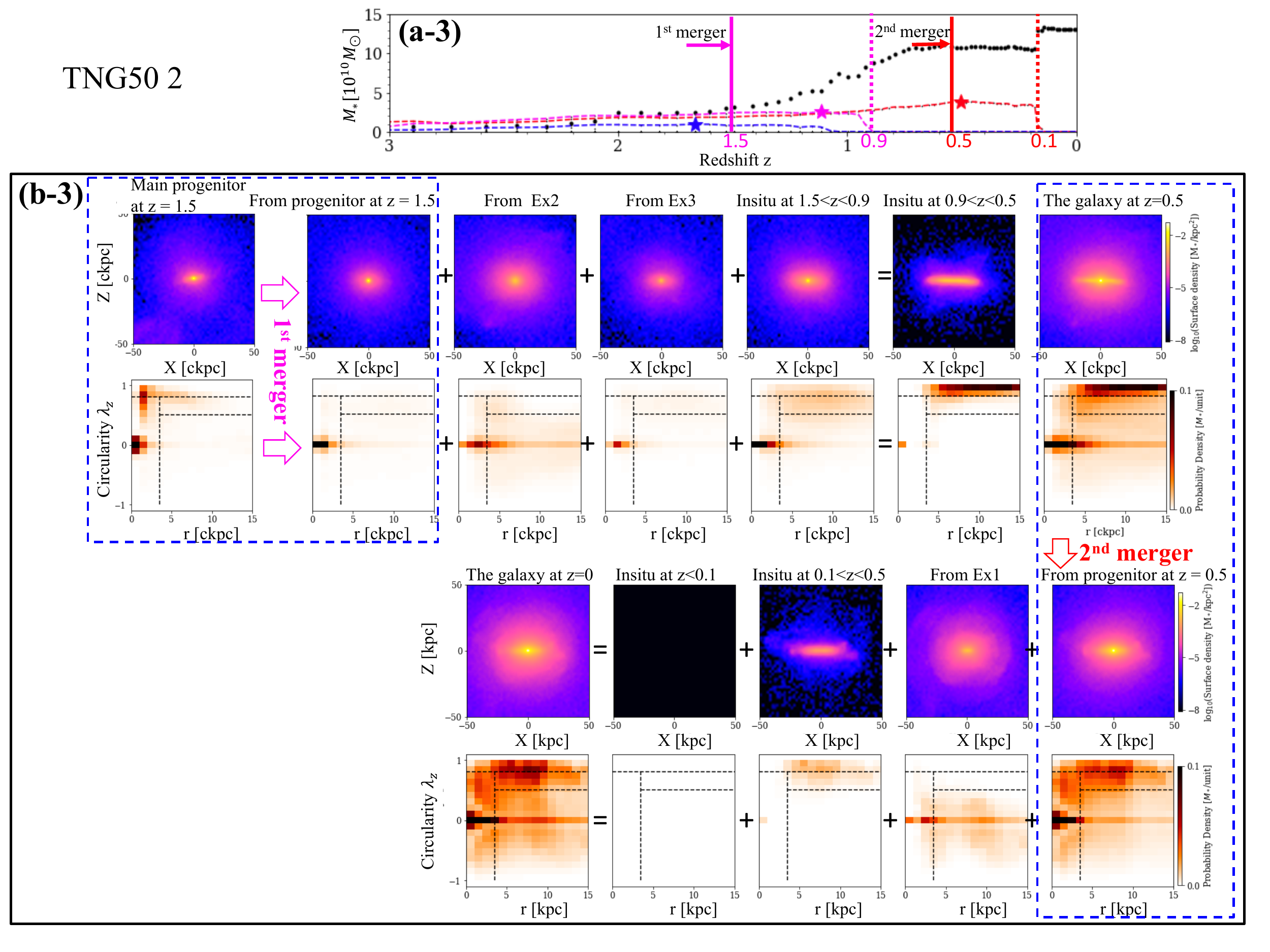}
\centering\includegraphics[width=14cm]{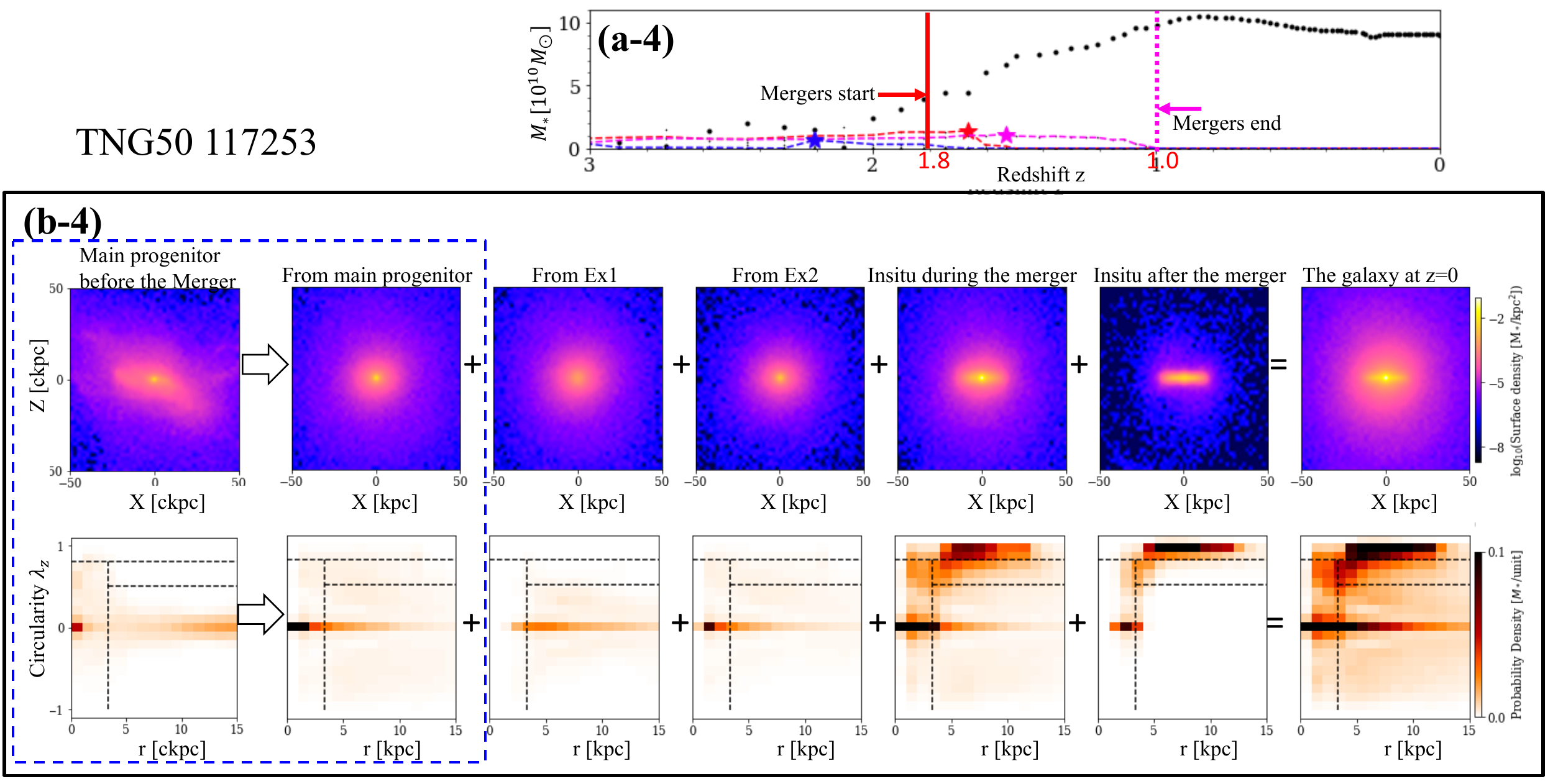}
\caption{
{Formation of the hot inner stellar halo in the simulated galaxies TNG50\,2 (a-3 and b-3) and TNG50\,117253 (a-4 and b-4).} The symbols are the same as in Fig.~\ref{fig:formation1}. 
{In panel  (a-3)}, TNG50\,2 has accreted three massive satellites, which are denoted by Ex1, Ex2, and Ex3, with decreasing stellar mass. Here, Ex3 and Ex2 were accreted simultaneously at $z\sim 1$ and marked as the first merger, and Ex3 was accreted at $z\sim0.1$ and marked as the second merger. {In panel  (b-3)} we trace the structure formation during the two merger events.  During the  first merger, stars accreted from both Ex2 and Ex3, as well as stars formed during the merger, contribute to the hot inner stellar halo.  After the first merger and before the second merger, a large number of stars were formed in a cold disk. The disk was then heated, becoming warm, during the second merger, and stars accreted from Ex1 contribute significantly to the hot inner stellar halo. Stars are rarely formed during and after the second merger as there is little gas left. TNG50\,2 has a hot inner stellar halo mass of $4.3\times10^{10}\,M_{\odot}$, of which $45\%$ is a product of the most massive merger. 
 {In panel  (a-4)} TNG50\, 117253 has accreted three
massive satellites, which are still denoted by Ex1, Ex2, and Ex3, with decreasing stellar mass.  We focus on the last two accretion events, Ex1 and Ex2, at 1 < z < 1.8.
In panel  (b-4)  we show the structure formation during the accretion of Ex1 and Ex2. TNG50 117253 has a hot inner stellar halo mass of $1.5\times10^{10}\,M_{\odot}$: the fraction produced by the most massive merger has an upper limit of $60\%$.
}
\label{fig:formation2}
\end{figure*}

We illustrate the formation of the hot inner stellar halos with four simulated galaxies from TNG50: namely, galaxies with Subhalo ID at $z=0$ reading 2, 468590, 117253, and 375073.  
These represent the high-mass end of the range under scrutiny, with galaxy stellar masses of $1.9, 1.4, 1.3, \rm$ and $ 1.9 \times10^{11}$\,\Msun, respectively. However, TNG50\,2 and 468590 are on the average mass-size relation, with stellar sizes, $R_e$, measuring 6.5 and 7.0 kpc, respectively -- they hence belong to subsample g6, as annotated in Figs.~\ref{fig:Macc_9} and \ref{fig:Macc3_9}. Galaxies TNG50\,117253 and 375073 lie below the average mass-size relation for their mass (grouping g9), with sizes of 3.1 and 5.2 kpc, respectively. They have a variety of
accretion histories, but all have a substantial hot inner stellar halo at $z=0$, between a few $10^{10}$ and $10^{11}$\,\Msun. In the following, we generally refer to ``massive mergers'' as to those whose secondary's stellar mass is of the same magnitude as the most massive one ever accreted (1:10); the rest are considered to be minor mergers.

The formation history of TNG50\,468590 is shown in panels (a-1) and (b-1) of Fig.~\ref{fig:formation1}.\ That of TNG50\,375073 is shown in panels (a-2) and (b-2).

The assembly history of TNG50\,468590 (top panel of Fig.~\ref{fig:formation1}) shows that this galaxy's past is dominated by a massive merger with a stellar mass ratio of about $1:1$ at redshift $z \sim 1$. This galaxy has accreted stellar mass for a total of $6\times 10^{10}$\,\Msun\, (ex situ fraction of $43\%$), with its most massive merger  contributing $94\%$ of it. In panel (b-1) we show the galactic structure of the progenitor of TNG50\,468590 before and after this merger. 
Before the merger, the main progenitor was dominated by a cold disk and a compact bulge with no substantial hot inner stellar halo.  During the merger, both the main progenitor and the satellite were destroyed, with some stars settling into the bulge and others forming part of the hot inner stellar halo. 
Both the main progenitor and the satellite were gas rich. As such, $4\times 10^{10}$\,\Msun\ of stars (comparable to the mass of the main progenitor at the time) formed during the merger, most of which contributed to the bulge and $\sim15\%$ were distributed into the hot inner stellar halo.

After the merger, new stars formed on dynamically cold disk-like orbits. This disk structure persisted until $z=0$ because the quiescent merger history after this event results in only little dynamical heating and a minor contribution of these ex situ stars to the hot inner stellar halo. The hot inner stellar halo mass of TNG50 468590 at $z=0$ is $3.4\times 10^{10}\,M_{\odot}$: importantly, only $\sim10\%$ of the inner-halo stars were already there in the main progenitor before the massive merger started; $\sim30\%$ were in the bulge and/or disk of the main progenitor and redistributed to the hot inner stellar halo during the merger; $\sim40\%$ were brought in by the satellite galaxy; and $20\%$ formed during the merger and were immediately distributed into the hot inner stellar halo regions. In total, about $90\%$ of the hot inner stellar halo of this simulated galaxy was the product of the massive merger event. 

Similar figures for TNG50\,375073 in panels (a-2) and (b-2) of Fig.~\ref{fig:formation1} show that TNG50\,375073 has undergone two massive mergers, one of which contributed $58\%$ of the total accreted stellar mass (i.e., $5.1\times 10^{10}$\,\Msun). This galaxy has a hot inner stellar halo mass of $3.0\times 10^{10}\,M_{\odot}$ at $z=0$. We also show in this case the galactic structure during the most massive merger, which started at  $z\sim 1.3$ with a stellar mass ratio of $\sim1:0.5$. At $z=1.3, $ before this merger started, $\sim 14\%$ of the hot inner stellar halo stars were already in the main progenitor, which might be the result of the less massive merger that happened earlier on. During the most massive merger, $\sim38\%$ of the stars of the $z=0$ inner halo were redistributed into the halo regions from the bulge and disk of the main progenitor; $\sim34\%$ were accreted from the satellite galaxy; and $\sim 12\%$ were formed during the merger event. In total, $84\%$ of the hot inner stellar halo mass of TNG50 375073 was induced by the most massive merger event.

The other two galaxies are characterized by a smaller fraction of the hot inner stellar halo mass produced by the most massive merger event.\ They are shown in Fig.~\ref{fig:formation2}.

The galaxy TNG50\,2 (panels (a-3) and (b-3) of Fig.~\ref{fig:formation2}) accreted a total of $9\times10^{10}$\,\Msun\ (ex situ fraction of $47\%$), with the most massive merger contributing $50\%$ of it. We split its merger history into two parts: $1.5<z<0.9$ when the second and third massive satellites (Ex2 and Ex3) were accreted, which we call the first merger, and $0.5<z<0.1$ when the accretion of the most massive merger (Ex1) happened, which we call the
second merger. This galaxy has a hot inner stellar halo mass at $z=0$ of $4.3\times 10^{10}\,M_{\odot}$, and only $5\%$ of it was already in place before the first merger. 

During the first merger, stars from Ex2 and from Ex3 are accreted to make up $29\%$ and $5\%$, respectively, of the final mass of the hot inner stellar halo, whereas stars formed during this merger event contribute $14\%$ of the hot inner stellar halo. After the first merger and before the second merger started, a cold disk was reformed. During the second merger, the satellite galaxy Ex1 deposited its stars, making up $40\%$ of the hot inner stellar halo. The cold disk was heated to be dynamically warm, with a small part of it heated to be dynamically hot; furthermore, a small amount of stars were formed during the merger and contribute to the inner stellar halo. The combination of these two amounts to $\sim 5\%$. In this case, in total only about $45\%$ of the hot inner stellar halo mass is a product of the most massive merger event.

Finally, in the case of TNG50\,117253 (panels (a-4) and (b-4) of Fig.~\ref{fig:formation2}), the total accreted stellar mass is $3.2\times10^{10}$\,\Msun, $35\%$ of which comes from its most massive merger. Three massive satellites were accreted in quick succession between $z=2.5$ and $z=1$. By considering a snapshot of this galaxy at $z=1.8$, when the first merger ended, we can see that $12\%$ of the hot inner stellar halo mass was already in place at that time. Then another two satellites interacted with the main progenitor simultaneously, and it is hard to distinguish the effects of these two satellites. During these merger events, we find that $\sim23\%$ of the hot inner stellar halo stars were accreted from the most massive merger (Ex1); $\sim15\%$ were accreted from the second massive satellite (Ex2); and $\sim40\%$ were formed and induced during the mergers. At the end, TNG50 117253 had a total hot inner stellar halo mass of $1.5\times10^{10}\,M_{\odot}$. The fraction produced by the most massive merger event is less than $60\%$. 

The four cases described above give a snapshot of the complex assembly history of galaxies and, in particular, of the formation processes of their hot inner stellar halo. Yet, in all four cases, the stars accreted from the most massive merger(s) favor highly radial orbits with $\lambda_z\sim 0$. This is consistent with massive satellites on radial orbits being the most likely to deposit stars in the inner regions \citep{Boylan2008}. Still, the hot inner stellar halo is not only a result of the stripping of ex situ stars from incoming satellites and mergers, however massive.

Our analysis of TNG50 galaxies reveals that the stars in the hot inner stellar halo of galaxies have three main origin channels: a) they can be accreted from the satellite(s); b) they can be kicked out from the bulge and/or disk of the main progenitor by a massive merger; or c) they can be formed during a massive merger. This is consistent with the ``dual'' halo found in the Milky Way \citep{Haywood2018, Grand2020} and previously theorized for Milky Way-like galaxies by a number of numerical simulations \citep[e.g.,][]{Zolotov2009, Pillepich2015}.

A galaxy could undergo a few massive mergers during its history, and all in principle can contribute to the formation of its hot inner stellar halo. Based on the four example galaxies studied above, we can conclude that, depending on a galaxy's merger and accretion history, its most massive merger event may directly (via ex situ stars) or indirectly (by inducing in situ star formation or in situ star displacement) contribute from $40\%$ to $90\%$ of the mass of the hot inner stellar halo. 

\begin{figure}
\centering\includegraphics[width=9cm]{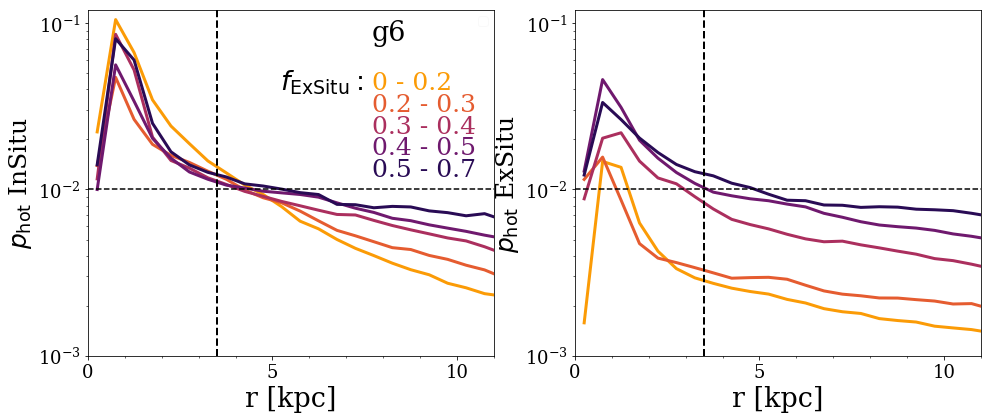}
\centering\includegraphics[width=8cm]{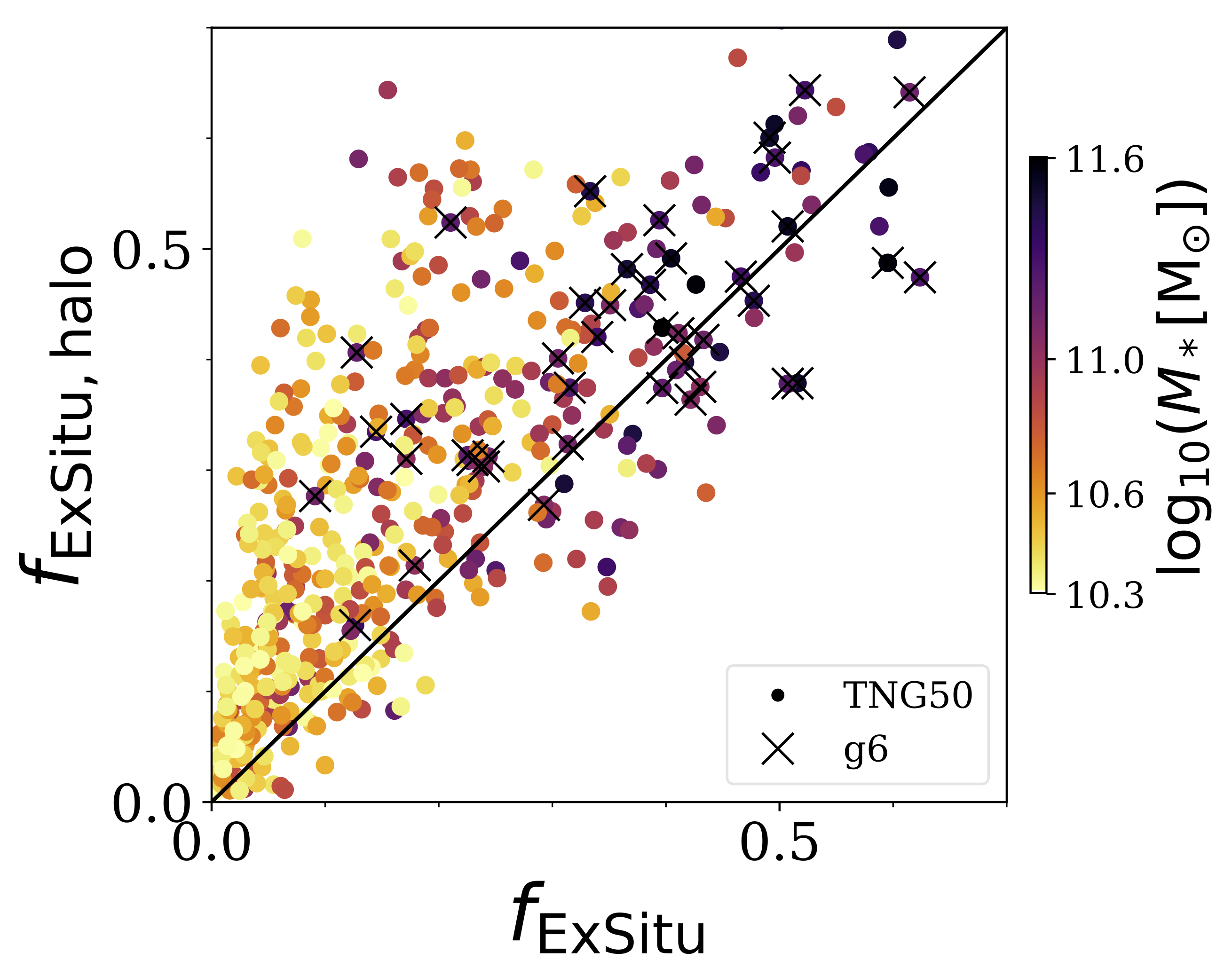}
\caption{{ Ex situ and in situ origin of the hot inner stellar halo of TNG50} galaxies. Top: Probability density distribution of stars on hot orbits, for in situ (left) and ex situ (right) stars separately. Here we divide the subsample of TNG50 galaxies in g6 into five bins according to their ex situ fraction, $f_{\mathrm{ExSitu}}$. Each curve represents the average $p_{\rm hot}$ (in situ or ex situ) of galaxies in each bin, color coded by its range of stellar ex situ fraction from 0-0.2 to 0.5-0.7, as indicated by the colored numbers. The curves are normalized such that the total probability of all stars within $2\,R_e$ equals unity. Bottom: Stellar ex situ fraction of the galaxy, $f_{\mathrm{ExSitu}}$, versus the ex situ fraction of stars in the hot inner stellar halo, $f_{\mathrm{ExSitu, halo}}$ ($3.5\,{\rm kpc}<r< 2R_e$ and $\lambda_z < 0.5$). Each dot represents one galaxy in TNG50, color coded by galaxies stellar mass, \Mstar, and crosses denote galaxies in subsample g6 (i.e., with a galaxy stellar mass in the range $10^{11}-10^{11.6}$ and an average stellar size).}
\label{fig:g6_InEx}
\end{figure}

\subsection{Contribution of in situ and ex situ stars}

From the study cases above, we have learned that the hot inner stellar halo as defined in Sect.~\ref{ss:components} is a mixture of in situ and ex situ stars. Here we extend the analysis above to all galaxies in the TNG50 sample. 
In the top panel of Fig.~\ref{fig:g6_InEx} we show the probability density distribution of stars on dynamically hot orbits as a function of radius, separately for in situ (left panel) and ex situ (right panel) stars. Here we further divide the subsample of galaxies in g6 into five bins according to their ex situ fraction. Each curve represents the average $p_{\rm hot}$ (in situ or ex situ) of galaxies in each bin, color coded by its range of stellar ex situ fraction. The curves are normalized such that the total probability of all stars within $2\,R_e$ equals unity. Thus, the galaxies with lower ex situ fractions have lower fractions of hot orbits, in situ and ex situ stars combined, due to higher fractions of disk orbits.

In galaxies with low ex situ fractions, the in situ hot stars are highly concentrated in the inner regions, with a low tail outside, whereas the ex situ stars are radially extended but with low density. In galaxies with higher ex situ fractions, the density tails of in situ stars at $r>3.5$ kpc are higher. Combining this with what we learned from the example galaxies above, this means that the distributions of in situ stars on previous bulges and/or disks are altered by the mergers, with more massive mergers inducing more prominent tails of the in situ hot stars at a large radius. At the same time, ex situ stars also contribute more to the tail with higher ex situ fractions, although about half of the ex situ stars sink into the center and become part of the bulge. In galaxies with stellar ex situ fractions $f_{\rm ExSitu} \gtrsim 0.5$, the density distributions of in situ and ex situ stars are similar, as they could have had major mergers with a mass ratio close to 1:1.

We note that a bulge with a similar size as that in galaxies with few ex situ stars formed after major mergers. The density of hot orbits peaks at a similar radius, with $\rmaxhot \sim 1$\, kpc for galaxies with different ex situ fractions.

In the bottom panel of Fig.~\ref{fig:g6_InEx} we further quantify the ex situ versus in situ origin of the hot inner stellar halo by analyzing all TNG50 galaxies in the selection. We show the stellar ex situ fraction of the whole galaxy, $f_{\rm ExSitu}$, versus the ex situ fraction of hot inner stellar halo, $f_{\rm ExSitu, halo}$. The hot inner stellar halos in general have a higher ex situ fraction than the whole galaxy, although the two values become close in the galaxies with higher (global) ex situ fractions. However, in most galaxies, particularly those with lower masses (i.e., lower ex situ fractions), the ex situ fraction of stars in the hot inner stellar halo is smaller than $50\%$\footnote{It should be kept in mind that here we focus on the inner stellar halo of galaxies, defined in this paper to be limited to within $2\,R_e$: the ex situ fraction is expected to be larger for dynamical hot orbits at larger radii \citep{Pillepich2018a}.}. 

All the findings presented above explain why a correlation naturally emerges between the mass of the most massive merger and the mass of the hot inner stellar halo of a galaxy, why such a correlation is stronger the higher the galaxy's stellar mass, and why, at the same time, the relationship exhibits a lingering non-negligible galaxy-to-galaxy variation. Determining the origin of such a variation is the next and final objective of the paper.

\subsection{Scatter of the correlations}
\subsubsection{Ex situ fraction of the hot inner stellar halo}
In Fig.~\ref{fig:Macc3_9} we show that there is a strong correlation between the mass of the hot inner stellar halo, \Mshalo, and the total ex situ stellar mass, $M_{*, {\rm Exsitu}}$. However, there is still scatter, especially in galaxies below the average mass-size relation (g4, g7, g8). This scatter could be related to the ex situ fraction of galaxies: how many stars are accreted, how much the galactic structures are altered, and how much star formation is triggered by the merger events.
As we have already seen in Fig.~\ref{fig:Macc3_tng50_1}, the outliers are mostly galaxies with $f_{\mathrm{ExSitu}}<0.1$, and the majority of galaxies with different ex situ fractions are closely distributed along the correlations.

\begin{figure}
\centering\includegraphics[width=9cm]{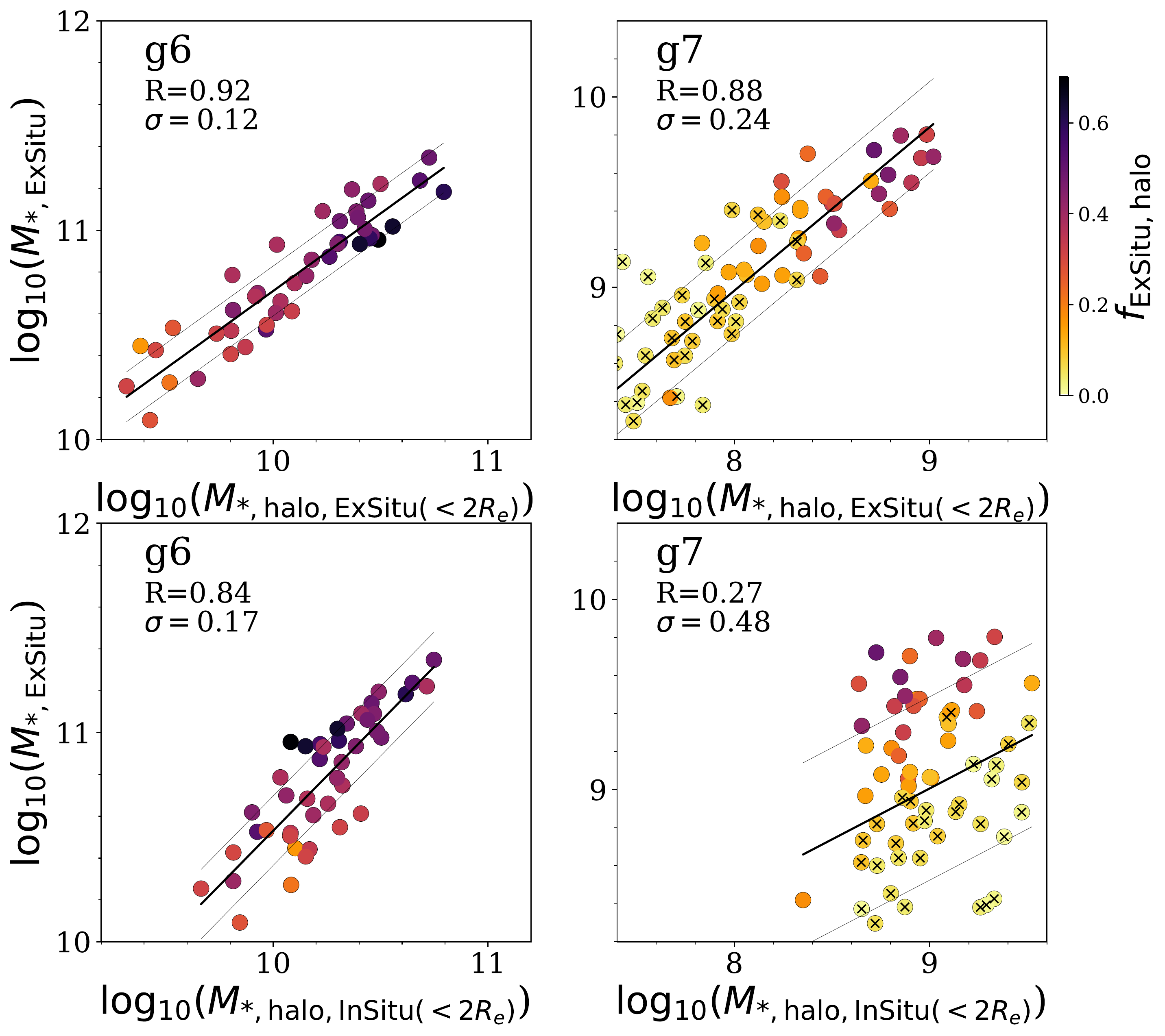}
\caption{{ Dependence of the relation between \Mshalo\, and $M_{*, {\rm ExSitu}}$ on the ex situ fraction of the hot inner stellar halo, $f_{\mathrm{ExSitu, halo}}$.}
We split the hot inner stellar halo into two parts: $M_{*,\mathrm{halo, ExSitu}}$ in the top panels and $M_{*,\mathrm{halo, InSitu}}$ in the bottom. Here we show results for TNG50 in subsamples g6 and g7, color coded by $f_{\mathrm{ExSitu, halo}}$.
The plus symbols mark the galaxies with $f_{\mathrm{ExSitu, halo}}<0.1$. In g6, both $M_{*,\mathrm{halo, ExSitu}}$ and $M_{*,\mathrm{halo, InSitu}}$ are tightly correlated with the total ex situ stellar mass.  In g7, $M_{*,\mathrm{halo, ExSitu}}$ shows a similarly strong correlation with the total ex situ stellar mass; however, the trend with $M_{*,\mathrm{halo, InSitu}}$ is weak with large scatter caused by the galaxies with low ex situ fractions of the hot inner stellar halo.
}
\label{fig:rt0}
\end{figure}

In order to understand the scatter, we separated the hot inner stellar halo into ex situ and in situ stars. In Fig.~\ref{fig:rt0} we show the correlation between the total ex situ stellar mass, $M_{*,{\rm ExSitu}}$, with $M_{*,\mathrm{halo, ExSitu}}$ (top) and $M_{*,\mathrm{halo, InSitu}}$ (bottom), respectively, for galaxies in g6 and g7.
As a reminder, we note that g6 is the group of galaxies with large masses ($10^{11}<M_*<10^{11.6}$), average sizes, $R_e$, and with a strong correlation of \Mshalo\, versus $M_{*, \mathrm{ExSitu}}$, as shown in Fig.~\ref{fig:Macc3_9}, while g7 is the group with small masses ($10^{10.3}<M_*<10^{10.6}$), small sizes ($R_e$ below the average), and with the weakest correlation of \Mshalo\, versus $M_{*, \mathrm{ExSitu}}$.

The data are color coded by the ex situ fraction of stars in the hot inner stellar halo, $f_{\rm ExSitu, halo}$, and not by the $f_{\mathrm{ExSitu}}$ of the galaxy.

In g6, both $M_{*,\mathrm{halo, ExSitu}}$ and $M_{*,\mathrm{halo, InSitu}}$ are tightly correlated with the total ex situ stellar mass, although the correlation is slightly weaker for the latter. 
The combination of these two components leads to the strong correlation of \Mshalo versus $M_{*,\mathrm{ExSitu}}$.
For these galaxies, mergers play the dominant role for the origin of both ex situ and in situ stars in the hot inner stellar halo. As we show in Sect.~\ref{ss:case}, in situ stars in the hot inner stellar halo should mainly be induced by mergers from a preexisting bulge and/or disk and formed during the star formation triggered by mergers. 

In g7, $M_{*,\mathrm{halo, ExSitu}}$ shows a similarly strong correlation with the total ex situ stellar mass.
However, the trend with $M_{*,\mathrm{halo, InSitu}}$ is weak. There are some galaxies with $f_{\mathrm{ExSitu, halo}}<0.1$, and, as we discuss in Sect.~\ref{sss:rcut} and show in Fig.~\ref{fig:rcdf_9}, there is no clear transition from bulge to hot inner stellar halo at $r\sim3.5$ kpc for galaxies with low ex situ fractions (i.e., with $f_{\mathrm{ExSitu}}<0.1$, including those with $f_{\mathrm{ExSitu, halo}}<0.1$). For these galaxies, the origin of the in situ stars in the hot inner stellar halo is largely independent of mergers and instead due to the intrinsic scattering of the density distribution of the bulge. As shown in the bottom-right panel of Fig.~\ref{fig:rt0}, all galaxies with $f_{\mathrm{ExSitu, halo}}<0.1$ lie below the correlation and cause the large scatter in $M_{*,\mathrm{halo, InSitu}}$ versus $M_{*,\mathrm{ExSitu}}$. The $M_{*,\mathrm{halo, InSitu}}$ is dominating the mass of the hot inner stellar halo, thus further causing the large scatter in the relation of \Mshalo versus $M_{*,\mathrm{ExSitu}}$.

\subsubsection{Dominance of the most massive merger}
In both the Milky Way and M31, the buildup of the inner stellar halo is thought to be dominated by one massive merger event \citep{Deason2015,Helmi2018, DSouza2018}. In more massive galaxies, the stellar halos are more likely to be built up through several large, equally massive mergers \citep{Cooper2013}. 
In our sample, which spans a wide mass range, the scatter of hot inner stellar halo mass, \Mshalo, versus the mass of the most massive merger, $M_{*, {\rm Ex1}}$, is further caused by the diversity of accretion histories and by whether and how much star formation is triggered during mergers. Here we show that the scatter is driven to a great extent by the number of massive merger events that have occurred over time. 

\begin{figure}
\centering\includegraphics[width=9cm]{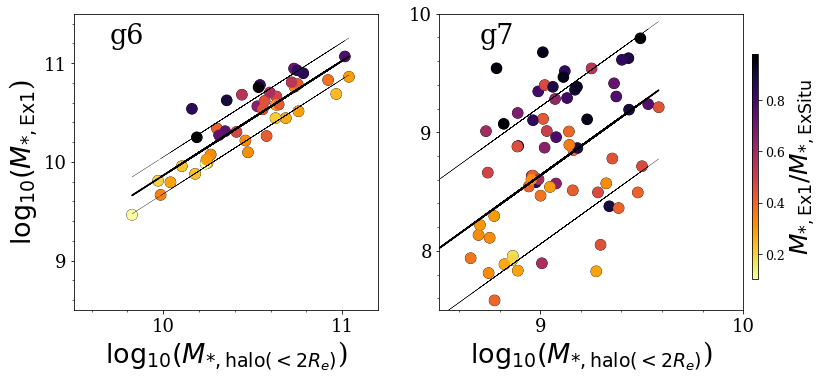}
\caption{{ Dependence of the relation between \Mshalo\, and $M_{*, {\rm Ex1}}$ on the dominance of the most massive merger event.} Here we show results for TNG50 in subsamples g6 and g7, color coded by the ratio between the ex situ stellar mass brought in by the most massive merger and the total ex situ stellar mass: $M_{*, {\rm Ex1}} / M_{*, {\rm ExSitu}}$. The larger this ratio is, the larger the contribution of the most massive merger to the formation of the hot inner stellar halo.
}
\label{fig:rt}
\end{figure}

In Fig.~\ref{fig:rt} we again show this correlation for galaxies in subsamples g6 and g7, but this time color coded by the contribution of the most massive merger to the total ex situ stellar mass: $M_{*, {\rm Ex1}} / M_{*, {\rm ExSitu}}$. Galaxies with accretion histories dominated by one massive merger tend to lie above the correlation: almost all the galaxies with
$M_{*, {\rm Ex1}} / M_{*, {\rm ExSitu}}>0.8$ lie above the median line.\ This is the case of TNG50\,468590, shown as one case study (see Fig.~\ref{fig:formation1}). On the other hand, galaxies with multiple massive mergers tend to lie below the median of the correlation, such as TNG50\,117253 (Fig.~\ref{fig:formation2}).

\section{Summary}

In this paper we have analyzed the cosmological galaxy simulations TNG50, TNG100, TNG300, and EAGLE at $z=0$. We find that the hot inner stellar halo of a galaxy, defined by the stars on dynamically hot orbits with $3.5\,{\rm kpc}<r< 2R_e$ and
$\lambda_z<0.5$, is a strong predictor of the total ever-accreted stellar mass and of the stellar mass accreted from the most massive merger the galaxy has ever experienced.

In particular, we find a strong correlation between the hot inner stellar halo mass of a galaxy, \Mshalo, and the total ex situ (i.e., accreted) stellar mass, $M_{\rm ExSitu}$. The analog correlation of \Mshalo\, versus the stellar mass of the most massive merger the galaxy has ever accreted, $M_{\rm Ex1}$, is also strong.

These correlations hold almost unchanged in simulated galaxies throughout the inspected galaxy mass range from TNG50 and TNG100 (\Mstar\, in the range of $10^{10.3-11.6}$\,\Msun), TNG300 (at $\Mstar>10^{11}$\,\Msun), and EAGLE (at $\Mstar>10^{10.6}$\,\Msun). Therefore, our results are robust against different galaxy formation (i.e., feedback) models and across varying numerical resolution. The main result holds for those galaxy formation models and resolution choices that reproduce to a reasonable level of degree the most fundamental observed galaxy properties and statistics, as is the case for the TNG and EAGLE simulations. We find that TNG300 galaxies at stellar masses $M_*\lesssim 10^{11}$\,\Msun\, and EAGLE galaxies at $M_*\lesssim 10^{10.6}$\,\Msun\, deviate toward larger \Mshalo\, at fixed satellite mass, and this is likely caused by an overestimation of \Mshalo\, due to their larger bulge sizes.

Using the outcome of TNG50, we also find that the correlation between the inner-halo mass and the total ex situ stellar mass (or mass of the most massive merger) is strongest for massive and extended galaxies with $M_*>10^{10.6}$\,\Msun\, and $R_e>3$\,kpc, with scatter on the order of $0.1-0.2$ dex; this scatter increases by an additional $0.1-0.3$ dex in galaxies of lower mass and smaller stellar size.  

The analysis of TNG50 galaxies clearly shows that the hot inner stellar halo is a product of the most massive merger(s) a galaxy has ever experienced. For galaxies with low ex situ stellar mass fractions, the radial probability density distributions of stars on dynamically hot
orbits are highly concentrated within $\sim3.5$\,kpc; on the other hand, galaxy mergers, and hence higher ex situ stellar mass fractions, produce
an extended component that contributes to the hot inner stellar halo. 
We find that, according to TNG50, the stars in the hot inner stellar halo can have
three main origins:
\begin{enumerate}
\item They can be accreted from satellite and merging galaxies.\\
\item They can be scattered out of the bulge and/or disk of a galaxy because of the interaction with satellite and merging galaxies.\\
\item They can be formed during the merger events and interactions.
\end{enumerate}
 The ex situ fraction within the hot inner stellar halo, origin pathway (1) of the inner-halo stars, is $\sim 50\%$ in the most massive galaxies but can be smaller in lower-mass galaxies. Hence, our analysis shows that the formation of the hot inner stellar halo is not only due to the accretion of ex situ stars from merging and satellite galaxies but is also driven by the dynamical processes that merger events can trigger, such as enhanced star formation episodes and displacement and heating of in situ stars.

The mass of the hot inner stellar halo of galaxies defined in this paper is a quantity that we can robustly obtain from observations, as we will show in a companion paper. In fact, the maximum extension we advocate to define the hot inner stellar halo in this paper, $2\,R_e$, is set by limitations of observational data coverage rather than physical motivations. Using the correlations quantified here we will show in a companion paper how we can infer the stellar mass of the most massive merger(s) ever accreted by galaxies in the nearby Universe.

\begin{acknowledgement}

We thank the useful discussion with people from the Fornax 3D team: Enrica Iodice, Francesca Pinna, Ignacio Martin Navarro, Lodovico Coccato,  Jes\'us Falc\'on-Barroso, Enrico Maria Corsini, Dimitri A. Gadotti, Katja Fahrion, Mariya Lyubenova, Richard McDermid, Adriano Poci, Mark Sarzi, and Tim de Zeeuw. LZhu acknowledges the support from the National Key
R$\&$D Program of China under grant No. 2018YFA0404501 and National Natural Science Foundation of China under grant No. Y945271001. AP acknowledges funding from the Deutsche Forschungsgemeinschaft (DFG, German Research Foundation) -- Project-ID 138713538 -- SFB 881 (``The Milky Way System'', subproject A01). GV acknowledges funding from the European Research Council (ERC) under the European Union's Horizon 2020 research and innovation programme under grant agreement No 724857 (Consolidator Grant ArcheoDyn). DN acknowledges funding from the Deutsche Forschungsgemeinschaft (DFG) through an Emmy Noether Research Group (grant number NE 2441/1-1).
TNG50 was realised with compute time granted by the Gauss Centre for Super-computing (GCS), under the GCS Large-Scale Project GCS-DWAR (2016; PIs Nelson/Pillepich).

\end{acknowledgement}

\bibliographystyle{aa}
\bibliography{ms_fcc167}

\begin{thebibliography}{90}
\expandafter\ifx\csname natexlab\endcsname\relax\def\natexlab#1{#1}\fi

\bibitem[{{Beasley} {et~al.}(2018){Beasley}, {Trujillo}, {Leaman}, \&
  {Montes}}]{Beasley2018}
{Beasley}, M.~A., {Trujillo}, I., {Leaman}, R., \& {Montes}, M. 2018, \nat,
  555, 483

\bibitem[{{Bell} {et~al.}(2017){Bell}, {Monachesi}, {Harmsen}, {de Jong},
  {Bailin}, {Radburn-Smith}, {D'Souza}, \& {Holwerda}}]{Bell2017}
{Bell}, E.~F., {Monachesi}, A., {Harmsen}, B., {et~al.} 2017, \apjl, 837, L8

\bibitem[{{Belokurov} {et~al.}(2018){Belokurov}, {Erkal}, {Evans}, {Koposov},
  \& {Deason}}]{Belokurov2018}
{Belokurov}, V., {Erkal}, D., {Evans}, N.~W., {Koposov}, S.~E., \& {Deason},
  A.~J. 2018, \mnras, 478, 611

\bibitem[{{Boecker} {et~al.}(2020){Boecker}, {Leaman}, {van de Ven}, {Norris},
  {Mackereth}, \& {Crain}}]{Boecker2020}
{Boecker}, A., {Leaman}, R., {van de Ven}, G., {et~al.} 2020, \mnras, 491, 823

\bibitem[{Bois {et~al.}(2010)Bois, Bournaud, Emsellem, Alatalo, Blitz, Bureau,
  Cappellari, Davies, Davis, de~Zeeuw, Duc, Khochfar, Krajnovi{\'{c}},
  Kuntschner, Lablanche, McDermid, Morganti, Naab, Oosterloo, Sarzi, Scott,
  Serra, Weijmans, \& Young}]{Bois2010}
Bois, M., Bournaud, F., Emsellem, E., {et~al.} 2010, Monthly Notices of the
  Royal Astronomical Society, 406, 2405

\bibitem[{Bois {et~al.}(2011)Bois, Emsellem, Bournaud, Alatalo, Blitz, Bureau,
  Cappellari, Davies, Davis, de~Zeeuw, Duc, Khochfar, Krajnovi{\'{c}},
  Kuntschner, Lablanche, McDermid, Morganti, Naab, Oosterloo, Sarzi, Scott,
  Serra, Weijmans, \& Young}]{Bois2011}
Bois, M., Emsellem, E., Bournaud, F., {et~al.} 2011, Monthly Notices of the
  Royal Astronomical Society, 416, 1654

\bibitem[{{Boylan-Kolchin} {et~al.}(2008){Boylan-Kolchin}, {Ma}, \&
  {Quataert}}]{Boylan2008}
{Boylan-Kolchin}, M., {Ma}, C.-P., \& {Quataert}, E. 2008, \mnras, 383, 93

\bibitem[{{Bundy} {et~al.}(2015){Bundy}, {Bershady}, {Law}, {Yan}, {Drory},
  {MacDonald}, {Wake}, {Cherinka}, {S{\'a}nchez-Gallego}, {Weijmans}, {Thomas},
  {Tremonti}, {Masters}, {Coccato}, {Diamond-Stanic}, {Arag{\'o}n-Salamanca},
  {Avila-Reese}, {Badenes}, {Falc{\'o}n-Barroso}, {Belfiore}, {Bizyaev},
  {Blanc}, {Bland-Hawthorn}, {Blanton}, {Brownstein}, {Byler}, {Cappellari},
  {Conroy}, {Dutton}, {Emsellem}, {Etherington}, {Frinchaboy}, {Fu}, {Gunn},
  {Harding}, {Johnston}, {Kauffmann}, {Kinemuchi}, {Klaene}, {Knapen},
  {Leauthaud}, {Li}, {Lin}, {Maiolino}, {Malanushenko}, {Malanushenko}, {Mao},
  {Maraston}, {McDermid}, {Merrifield}, {Nichol}, {Oravetz}, {Pan}, {Parejko},
  {Sanchez}, {Schlegel}, {Simmons}, {Steele}, {Steinmetz}, {Thanjavur},
  {Thompson}, {Tinker}, {van den Bosch}, {Westfall}, {Wilkinson}, {Wright},
  {Xiao}, \& {Zhang}}]{Bundy2015}
{Bundy}, K., {Bershady}, M.~A., {Law}, D.~R., {et~al.} 2015, \apj, 798, 7

\bibitem[{{Cappellari} {et~al.}(2007){Cappellari}, {Emsellem}, {Bacon},
  {Bureau}, {Davies}, {de Zeeuw}, {Falc{\'o}n-Barroso}, {Krajnovi{\'c}},
  {Kuntschner}, {McDermid}, {Peletier}, {Sarzi}, {van den Bosch}, \& {van de
  Ven}}]{Cappellari2007}
{Cappellari}, M., {Emsellem}, E., {Bacon}, R., {et~al.} 2007, \mnras, 379, 418

\bibitem[{{Cappellari} {et~al.}(2011){Cappellari}, {Emsellem}, {Krajnovi{\'c}},
  {McDermid}, {Scott}, {Verdoes Kleijn}, {Young}, {Alatalo}, {Bacon}, {Blitz},
  {Bois}, {Bournaud}, {Bureau}, {Davies}, {Davis}, {de Zeeuw}, {Duc},
  {Khochfar}, {Kuntschner}, {Lablanche}, {Morganti}, {Naab}, {Oosterloo},
  {Sarzi}, {Serra}, \& {Weijmans}}]{Cappellari2011}
{Cappellari}, M., {Emsellem}, E., {Krajnovi{\'c}}, D., {et~al.} 2011, \mnras,
  413, 813

\bibitem[{{Cooper} {et~al.}(2013){Cooper}, {D'Souza}, {Kauffmann}, {Wang},
  {Boylan-Kolchin}, {Guo}, {Frenk}, \& {White}}]{Cooper2013}
{Cooper}, A.~P., {D'Souza}, R., {Kauffmann}, G., {et~al.} 2013, \mnras, 434,
  3348

\bibitem[{{Correa} {et~al.}(2017){Correa}, {Schaye}, {Clauwens}, {Bower},
  {Crain}, {Schaller}, {Theuns}, \& {Thob}}]{Correa2017}
{Correa}, C.~A., {Schaye}, J., {Clauwens}, B., {et~al.} 2017, \mnras, 472, L45

\bibitem[{Cox {et~al.}(2006)Cox, Dutta, Matteo, Hernquist, Hopkins, Robertson,
  \& Springel}]{Cox2006}
Cox, T.~J., Dutta, S.~N., Matteo, T.~D., {et~al.} 2006, The Astrophysical
  Journal, 650, 791

\bibitem[{{Crain} {et~al.}(2015){Crain}, {Schaye}, {Bower}, {Furlong},
  {Schaller}, {Theuns}, {Dalla Vecchia}, {Frenk}, {McCarthy}, {Helly},
  {Jenkins}, {Rosas-Guevara}, {White}, \& {Trayford}}]{Crain2015}
{Crain}, R.~A., {Schaye}, J., {Bower}, R.~G., {et~al.} 2015, \mnras, 450, 1937

\bibitem[{{Croom} {et~al.}(2012){Croom}, {Lawrence}, {Bland-Hawthorn},
  {Bryant}, {Fogarty}, {Richards}, {Goodwin}, {Farrell}, {Miziarski}, {Heald},
  {Jones}, {Lee}, {Colless}, {Brough}, {Hopkins}, {Bauer}, {Birchall}, {Ellis},
  {Horton}, {Leon-Saval}, {Lewis}, {L{\'o}pez-S{\'a}nchez}, {Min}, {Trinh}, \&
  {Trowland}}]{Croom2012}
{Croom}, S.~M., {Lawrence}, J.~S., {Bland-Hawthorn}, J., {et~al.} 2012, \mnras,
  421, 872

\bibitem[{{Davies} {et~al.}(2001){Davies}, {Kuntschner}, {Emsellem}, {Bacon},
  {Bureau}, {Carollo}, {Copin}, {Miller}, {Monnet}, {Peletier}, {Verolme}, \&
  {de Zeeuw}}]{Davies2001}
{Davies}, R.~L., {Kuntschner}, H., {Emsellem}, E., {et~al.} 2001, \apjl, 548,
  L33

\bibitem[{{Davison} {et~al.}(2021){Davison}, {Norris}, {Leaman}, {Kuntschner},
  {Boecker}, \& {van de Ven}}]{Davison2021}
{Davison}, T.~A., {Norris}, M.~A., {Leaman}, R., {et~al.} 2021, \mnras, 507,
  3089

\bibitem[{{Davison} {et~al.}(2020){Davison}, {Norris}, {Pfeffer}, {Davies}, \&
  {Crain}}]{Davison2020}
{Davison}, T.~A., {Norris}, M.~A., {Pfeffer}, J.~L., {Davies}, J.~J., \&
  {Crain}, R.~A. 2020, \mnras, 497, 81

\bibitem[{{de Graaff} {et~al.}(2021){de Graaff}, {Trayford}, {Franx},
  {Schaller}, {Schaye}, \& {van der Wel}}]{deGraaff2021}
{de Graaff}, A., {Trayford}, J., {Franx}, M., {et~al.} 2021, \mnras
  [\eprint[arXiv]{2110.02235}]

\bibitem[{{Deason} {et~al.}(2015){Deason}, {Belokurov}, \&
  {Weisz}}]{Deason2015}
{Deason}, A.~J., {Belokurov}, V., \& {Weisz}, D.~R. 2015, \mnras, 448, L77

\bibitem[{{Dolag} {et~al.}(2009){Dolag}, {Borgani}, {Murante}, \&
  {Springel}}]{Dolag2009}
{Dolag}, K., {Borgani}, S., {Murante}, G., \& {Springel}, V. 2009, \mnras, 399,
  497

\bibitem[{{D'Souza} \& {Bell}(2018{\natexlab{a}})}]{DSouza2018}
{D'Souza}, R. \& {Bell}, E.~F. 2018{\natexlab{a}}, Nature Astronomy, 2, 737

\bibitem[{{D'Souza} \& {Bell}(2018{\natexlab{b}})}]{DSouza2018b}
{D'Souza}, R. \& {Bell}, E.~F. 2018{\natexlab{b}}, \mnras, 474, 5300

\bibitem[{{Du} {et~al.}(2021){Du}, {Ho}, {Debattista}, {Pillepich}, {Nelson},
  {Hernquist}, \& {Weinberger}}]{Du2021}
{Du}, M., {Ho}, L.~C., {Debattista}, V.~P., {et~al.} 2021, \apj, 919, 135

\bibitem[{{Du} {et~al.}(2020){Du}, {Ho}, {Debattista}, {Pillepich}, {Nelson},
  {Zhao}, \& {Hernquist}}]{Du2020}
{Du}, M., {Ho}, L.~C., {Debattista}, V.~P., {et~al.} 2020, \apj, 895, 139

\bibitem[{{Du} {et~al.}(2019){Du}, {Ho}, {Zhao}, {Shi}, {Debattista},
  {Hernquist}, \& {Nelson}}]{Du2019}
{Du}, M., {Ho}, L.~C., {Zhao}, D., {et~al.} 2019, \apj, 884, 129

\bibitem[{{Fall} \& {Efstathiou}(1980)}]{Fall1980}
{Fall}, S.~M. \& {Efstathiou}, G. 1980, \mnras, 193, 189

\bibitem[{{Forbes} {et~al.}(2016){Forbes}, {Romanowsky}, {Pastorello},
  {Foster}, {Brodie}, {Strader}, {Usher}, \& {Pota}}]{Forbes2016}
{Forbes}, D.~A., {Romanowsky}, A.~J., {Pastorello}, N., {et~al.} 2016, \mnras,
  457, 1242

\bibitem[{{Genel} {et~al.}(2015){Genel}, {Fall}, {Hernquist}, {Vogelsberger},
  {Snyder}, {Rodriguez-Gomez}, {Sijacki}, \& {Springel}}]{Genel2015}
{Genel}, S., {Fall}, S.~M., {Hernquist}, L., {et~al.} 2015, \apjl, 804, L40

\bibitem[{{Genel} {et~al.}(2018){Genel}, {Nelson}, {Pillepich}, {Springel},
  {Pakmor}, {Weinberger}, {Hernquist}, {Naiman}, {Vogelsberger}, {Marinacci},
  \& {Torrey}}]{Genel2018}
{Genel}, S., {Nelson}, D., {Pillepich}, A., {et~al.} 2018, \mnras, 474, 3976

\bibitem[{Grand {et~al.}(2020)Grand, Kawata, Belokurov, Deason, Fattahi,
  Fragkoudi, G{\'{o}}mez, Marinacci, \& Pakmor}]{Grand2020}
Grand, R. J.~J., Kawata, D., Belokurov, V., {et~al.} 2020, Monthly Notices of
  the Royal Astronomical Society, 497, 1603

\bibitem[{{Harmsen} {et~al.}(2017){Harmsen}, {Monachesi}, {Bell}, {de Jong},
  {Bailin}, {Radburn-Smith}, \& {Holwerda}}]{Harmsen2017}
{Harmsen}, B., {Monachesi}, A., {Bell}, E.~F., {et~al.} 2017, \mnras, 466, 1491

\bibitem[{{Haywood} {et~al.}(2018){Haywood}, {Di Matteo}, {Lehnert}, {Snaith},
  {Khoperskov}, \& {G{\'o}mez}}]{Haywood2018}
{Haywood}, M., {Di Matteo}, P., {Lehnert}, M.~D., {et~al.} 2018, \apj, 863, 113

\bibitem[{{Helmi}(2020)}]{Helmi2020}
{Helmi}, A. 2020, \araa, 58, 205

\bibitem[{{Helmi} {et~al.}(2018){Helmi}, {Babusiaux}, {Koppelman}, {Massari},
  {Veljanoski}, \& {Brown}}]{Helmi2018}
{Helmi}, A., {Babusiaux}, C., {Koppelman}, H.~H., {et~al.} 2018, \nat, 563, 85

\bibitem[{Hoffman {et~al.}(2010)Hoffman, Cox, Dutta, \&
  Hernquist}]{Hoffman2010}
Hoffman, L., Cox, T.~J., Dutta, S., \& Hernquist, L. 2010, The Astrophysical
  Journal, 723, 818

\bibitem[{{Jin} {et~al.}(2020){Jin}, {Zhu}, {Long}, {Mao}, {Wang}, \& {van de
  Ven}}]{Jin2020}
{Jin}, Y., {Zhu}, L., {Long}, R.~J., {et~al.} 2020, \mnras, 491, 1690

\bibitem[{{Kormendy} \& {Kennicutt}(2004)}]{Kormendy2004}
{Kormendy}, J. \& {Kennicutt}, Robert~C., J. 2004, \araa, 42, 603

\bibitem[{{Kruijssen} {et~al.}(2019){Kruijssen}, {Pfeffer}, {Reina-Campos},
  {Crain}, \& {Bastian}}]{Kruijssen2019}
{Kruijssen}, J.~M.~D., {Pfeffer}, J.~L., {Reina-Campos}, M., {Crain}, R.~A., \&
  {Bastian}, N. 2019, \mnras, 486, 3180

\bibitem[{{Lagos} {et~al.}(2018){Lagos}, {Stevens}, {Bower}, {Davis},
  {Contreras}, {Padilla}, {Obreschkow}, {Croton}, {Trayford}, {Welker}, \&
  {Theuns}}]{Lagos2018}
{Lagos}, C. d.~P., {Stevens}, A. R.~H., {Bower}, R.~G., {et~al.} 2018, \mnras,
  473, 4956

\bibitem[{{Lange} {et~al.}(2016){Lange}, {Moffett}, {Driver}, {Robotham},
  {Lagos}, {Kelvin}, {Conselice}, {Margalef-Bentabol}, {Alpaslan}, {Baldry},
  {Bland-Hawthorn}, {Bremer}, {Brough}, {Cluver}, {Colless}, {Davies},
  {H{\"a}u{\ss}ler}, {Holwerda}, {Hopkins}, {Kafle}, {Kennedy}, {Liske},
  {Phillipps}, {Popescu}, {Taylor}, {Tuffs}, {van Kampen}, \&
  {Wright}}]{Lange2016}
{Lange}, R., {Moffett}, A.~J., {Driver}, S.~P., {et~al.} 2016, \mnras, 462,
  1470

\bibitem[{{Marinacci} {et~al.}(2018){Marinacci}, {Vogelsberger}, {Pakmor},
  {Torrey}, {Springel}, {Hernquist}, {Nelson}, {Weinberger}, {Pillepich},
  {Naiman}, \& {Genel}}]{Marinacci2018}
{Marinacci}, F., {Vogelsberger}, M., {Pakmor}, R., {et~al.} 2018, \mnras, 480,
  5113

\bibitem[{{Martig} {et~al.}(2021){Martig}, {Pinna}, {Falc{\'o}n-Barroso},
  {Gadotti}, {Husemann}, {Minchev}, {Neumann}, {Ruiz-Lara}, \& {van de
  Ven}}]{Martig2021}
{Martig}, M., {Pinna}, F., {Falc{\'o}n-Barroso}, J., {et~al.} 2021, \mnras,
  508, 2458

\bibitem[{{McAlpine} {et~al.}(2016){McAlpine}, {Helly}, {Schaller}, {Trayford},
  {Qu}, {Furlong}, {Bower}, {Crain}, {Schaye}, {Theuns}, {Dalla Vecchia},
  {Frenk}, {McCarthy}, {Jenkins}, {Rosas-Guevara}, {White}, {Baes}, {Camps}, \&
  {Lemson}}]{McAlpine2016}
{McAlpine}, S., {Helly}, J.~C., {Schaller}, M., {et~al.} 2016, Astronomy and
  Computing, 15, 72

\bibitem[{{Merritt} {et~al.}(2016){Merritt}, {van Dokkum}, {Abraham}, \&
  {Zhang}}]{Merritt2016}
{Merritt}, A., {van Dokkum}, P., {Abraham}, R., \& {Zhang}, J. 2016, \apj, 830,
  62

\bibitem[{{Naab} {et~al.}(2009){Naab}, {Johansson}, \& {Ostriker}}]{Naab2009}
{Naab}, T., {Johansson}, P.~H., \& {Ostriker}, J.~P. 2009, \apjl, 699, L178

\bibitem[{Naab {et~al.}(2014)Naab, Oser, Emsellem, Cappellari, Krajnovi{\'{c}},
  McDermid, Alatalo, Bayet, Blitz, Bois, Bournaud, Bureau, Crocker, Davies,
  Davis, de~Zeeuw, Duc, Hirschmann, Johansson, Khochfar, Kuntschner, Morganti,
  Oosterloo, Sarzi, Scott, Serra, van~de Ven, Weijmans, \& Young}]{Naab2014}
Naab, T., Oser, L., Emsellem, E., {et~al.} 2014, Monthly Notices of the Royal
  Astronomical Society, 444, 3357

\bibitem[{{Naiman} {et~al.}(2018){Naiman}, {Pillepich}, {Springel},
  {Ramirez-Ruiz}, {Torrey}, {Vogelsberger}, {Pakmor}, {Nelson}, {Marinacci},
  {Hernquist}, {Weinberger}, \& {Genel}}]{Naiman2018}
{Naiman}, J.~P., {Pillepich}, A., {Springel}, V., {et~al.} 2018, \mnras, 477,
  1206

\bibitem[{{Nelson} {et~al.}(2015){Nelson}, {Pillepich}, {Genel},
  {Vogelsberger}, {Springel}, {Torrey}, {Rodriguez-Gomez}, {Sijacki}, {Snyder},
  {Griffen}, {Marinacci}, {Blecha}, {Sales}, {Xu}, \& {Hernquist}}]{Nelson2015}
{Nelson}, D., {Pillepich}, A., {Genel}, S., {et~al.} 2015, Astronomy and
  Computing, 13, 12

\bibitem[{{Nelson} {et~al.}(2019{\natexlab{a}}){Nelson}, {Pillepich},
  {Springel}, {Pakmor}, {Weinberger}, {Genel}, {Torrey}, {Vogelsberger},
  {Marinacci}, \& {Hernquist}}]{Nelson2019}
{Nelson}, D., {Pillepich}, A., {Springel}, V., {et~al.} 2019{\natexlab{a}},
  \mnras, 490, 3234

\bibitem[{{Nelson} {et~al.}(2018){Nelson}, {Pillepich}, {Springel},
  {Weinberger}, {Hernquist}, {Pakmor}, {Genel}, {Torrey}, {Vogelsberger},
  {Kauffmann}, {Marinacci}, \& {Naiman}}]{Nelson2018}
{Nelson}, D., {Pillepich}, A., {Springel}, V., {et~al.} 2018, \mnras, 475, 624

\bibitem[{{Nelson} {et~al.}(2019{\natexlab{b}}){Nelson}, {Springel},
  {Pillepich}, {Rodriguez-Gomez}, {Torrey}, {Genel}, {Vogelsberger}, {Pakmor},
  {Marinacci}, {Weinberger}, {Kelley}, {Lovell}, {Diemer}, \&
  {Hernquist}}]{Nelson2019release}
{Nelson}, D., {Springel}, V., {Pillepich}, A., {et~al.} 2019{\natexlab{b}},
  Computational Astrophysics and Cosmology, 6, 2

\bibitem[{{Nelson} {et~al.}(2021){Nelson}, {Tacchella}, {Diemer}, {Leja},
  {Hernquist}, {Whitaker}, {Weinberger}, {Pillepich}, {Nelson}, {Terrazas},
  {Nevin}, {Brammer}, {Burkhart}, {Cochrane}, {van Dokkum}, {Johnson},
  {Marinacci}, {Mowla}, {Pakmor}, {Skelton}, {Speagle}, {Springel}, {Torrey},
  {Vogelsberger}, \& {Wuyts}}]{ENelson2021}
{Nelson}, E.~J., {Tacchella}, S., {Diemer}, B., {et~al.} 2021, \mnras, 508, 219

\bibitem[{{Obreja} {et~al.}(2019){Obreja}, {Dutton}, {Macci{\`o}}, {Moster},
  {Buck}, {van de Ven}, {Wang}, {Stinson}, \& {Zhu}}]{Obreja2019}
{Obreja}, A., {Dutton}, A.~A., {Macci{\`o}}, A.~V., {et~al.} 2019, \mnras, 487,
  4424

\bibitem[{{Pillepich} {et~al.}(2015){Pillepich}, {Madau}, \&
  {Mayer}}]{Pillepich2015}
{Pillepich}, A., {Madau}, P., \& {Mayer}, L. 2015, \apj, 799, 184

\bibitem[{{Pillepich} {et~al.}(2018{\natexlab{a}}){Pillepich}, {Nelson},
  {Hernquist}, {Springel}, {Pakmor}, {Torrey}, {Weinberger}, {Genel}, {Naiman},
  {Marinacci}, \& {Vogelsberger}}]{Pillepich2018b}
{Pillepich}, A., {Nelson}, D., {Hernquist}, L., {et~al.} 2018{\natexlab{a}},
  \mnras, 475, 648

\bibitem[{{Pillepich} {et~al.}(2019){Pillepich}, {Nelson}, {Springel},
  {Pakmor}, {Torrey}, {Weinberger}, {Vogelsberger}, {Marinacci}, {Genel}, {van
  der Wel}, \& {Hernquist}}]{Pillepich2019}
{Pillepich}, A., {Nelson}, D., {Springel}, V., {et~al.} 2019, \mnras, 490, 3196

\bibitem[{{Pillepich} {et~al.}(2018{\natexlab{b}}){Pillepich}, {Springel},
  {Nelson}, {Genel}, {Naiman}, {Pakmor}, {Hernquist}, {Torrey}, {Vogelsberger},
  {Weinberger}, \& {Marinacci}}]{Pillepich2018a}
{Pillepich}, A., {Springel}, V., {Nelson}, D., {et~al.} 2018{\natexlab{b}},
  \mnras, 473, 4077

\bibitem[{{Pinna} {et~al.}(2019){Pinna}, {Falc{\'o}n-Barroso}, {Martig},
  {Sarzi}, {Coccato}, {Iodice}, {Corsini}, {de Zeeuw}, {Gadotti}, {Leaman},
  {Lyubenova}, {McDermid}, {Minchev}, {Morelli}, {van de Ven}, \&
  {Viaene}}]{Pinna2019}
{Pinna}, F., {Falc{\'o}n-Barroso}, J., {Martig}, M., {et~al.} 2019, \aap, 623,
  A19

\bibitem[{{Poci} {et~al.}(2021){Poci}, {McDermid}, {Lyubenova}, {Zhu}, {van de
  Ven}, {Iodice}, {Coccato}, {Pinna}, {Corsini}, {Falc{\'o}n-Barroso},
  {Gadotti}, {Grand}, {Fahrion}, {Mart{\'\i}n-Navarro}, {Sarzi}, {Viaene}, \&
  {de Zeeuw}}]{Poci2021}
{Poci}, A., {McDermid}, R.~M., {Lyubenova}, M., {et~al.} 2021, \aap, 647, A145

\bibitem[{{Poci} {et~al.}(2019){Poci}, {McDermid}, {Zhu}, \& {van de
  Ven}}]{Poci2019}
{Poci}, A., {McDermid}, R.~M., {Zhu}, L., \& {van de Ven}, G. 2019, \mnras,
  487, 3776

\bibitem[{{Pulsoni} {et~al.}(2020){Pulsoni}, {Gerhard}, {Arnaboldi},
  {Pillepich}, {Nelson}, {Hernquist}, \& {Springel}}]{Pulsoni2020}
{Pulsoni}, C., {Gerhard}, O., {Arnaboldi}, M., {et~al.} 2020, \aap, 641, A60

\bibitem[{{Pulsoni} {et~al.}(2021){Pulsoni}, {Gerhard}, {Arnaboldi},
  {Pillepich}, {Rodriguez-Gomez}, {Nelson}, {Hernquist}, \&
  {Springel}}]{Pulsoni2021}
{Pulsoni}, C., {Gerhard}, O., {Arnaboldi}, M., {et~al.} 2021, \aap, 647, A95

\bibitem[{{Rejkuba} {et~al.}(2011){Rejkuba}, {Harris}, {Greggio}, \&
  {Harris}}]{Rejkuba2011}
{Rejkuba}, M., {Harris}, W.~E., {Greggio}, L., \& {Harris}, G.~L.~H. 2011,
  \aap, 526, A123

\bibitem[{{Rodriguez-Gomez} {et~al.}(2015){Rodriguez-Gomez}, {Genel},
  {Vogelsberger}, {Sijacki}, {Pillepich}, {Sales}, {Torrey}, {Snyder},
  {Nelson}, {Springel}, {Ma}, \& {Hernquist}}]{Rodriguez2015}
{Rodriguez-Gomez}, V., {Genel}, S., {Vogelsberger}, M., {et~al.} 2015, \mnras,
  449, 49

\bibitem[{{Rodriguez-Gomez} {et~al.}(2016){Rodriguez-Gomez}, {Pillepich},
  {Sales}, {Genel}, {Vogelsberger}, {Zhu}, {Wellons}, {Nelson}, {Torrey},
  {Springel}, {Ma}, \& {Hernquist}}]{Rodriguez2016}
{Rodriguez-Gomez}, V., {Pillepich}, A., {Sales}, L.~V., {et~al.} 2016, \mnras,
  458, 2371

\bibitem[{{Rodriguez-Gomez} {et~al.}(2019){Rodriguez-Gomez}, {Snyder}, {Lotz},
  {Nelson}, {Pillepich}, {Springel}, {Genel}, {Weinberger}, {Tacchella},
  {Pakmor}, {Torrey}, {Marinacci}, {Vogelsberger}, {Hernquist}, \&
  {Thilker}}]{Rodriguez-Gomez2019}
{Rodriguez-Gomez}, V., {Snyder}, G.~F., {Lotz}, J.~M., {et~al.} 2019, \mnras,
  483, 4140

\bibitem[{{Sarzi} {et~al.}(2018){Sarzi}, {Iodice}, {Coccato}, {Corsini}, {de
  Zeeuw}, {Falc{\'o}n-Barroso}, {Gadotti}, {Lyubenova}, {McDermid}, {van de
  Ven}, {Fahrion}, {Pizzella}, \& {Zhu}}]{Sarzi2018}
{Sarzi}, M., {Iodice}, E., {Coccato}, L., {et~al.} 2018, \aap, 616, A121

\bibitem[{{Schaye} {et~al.}(2015){Schaye}, {Crain}, {Bower}, {Furlong},
  {Schaller}, {Theuns}, {Dalla Vecchia}, {Frenk}, {McCarthy}, {Helly},
  {Jenkins}, {Rosas-Guevara}, {White}, {Baes}, {Booth}, {Camps}, {Navarro},
  {Qu}, {Rahmati}, {Sawala}, {Thomas}, \& {Trayford}}]{Schaye2015}
{Schaye}, J., {Crain}, R.~A., {Bower}, R.~G., {et~al.} 2015, \mnras, 446, 521

\bibitem[{{Shen} {et~al.}(2003){Shen}, {Mo}, {White}, {Blanton}, {Kauffmann},
  {Voges}, {Brinkmann}, \& {Csabai}}]{Shen2003}
{Shen}, S., {Mo}, H.~J., {White}, S. D.~M., {et~al.} 2003, \mnras, 343, 978

\bibitem[{{Spavone} {et~al.}(2020){Spavone}, {Iodice}, {van de Ven},
  {Falc{\'o}n-Barroso}, {Raj}, {Hilker}, {Peletier}, {Capaccioli}, {Mieske},
  {Venhola}, {Napolitano}, {Cantiello}, {Paolillo}, \&
  {Schipani}}]{Spavone2020}
{Spavone}, M., {Iodice}, E., {van de Ven}, G., {et~al.} 2020, \aap, 639, A14

\bibitem[{{Springel} {et~al.}(2018){Springel}, {Pakmor}, {Pillepich},
  {Weinberger}, {Nelson}, {Hernquist}, {Vogelsberger}, {Genel}, {Torrey},
  {Marinacci}, \& {Naiman}}]{Springel2018}
{Springel}, V., {Pakmor}, R., {Pillepich}, A., {et~al.} 2018, \mnras, 475, 676

\bibitem[{{Springel} {et~al.}(2001){Springel}, {White}, {Tormen}, \&
  {Kauffmann}}]{Springel2001}
{Springel}, V., {White}, S. D.~M., {Tormen}, G., \& {Kauffmann}, G. 2001,
  \mnras, 328, 726

\bibitem[{{Trayford} {et~al.}(2019){Trayford}, {Frenk}, {Theuns}, {Schaye}, \&
  {Correa}}]{Trayford2019}
{Trayford}, J.~W., {Frenk}, C.~S., {Theuns}, T., {Schaye}, J., \& {Correa}, C.
  2019, \mnras, 483, 744

\bibitem[{{van den Bosch} {et~al.}(2008){van den Bosch}, {van de Ven},
  {Verolme}, {Cappellari}, \& {de Zeeuw}}]{vdB2008}
{van den Bosch}, R.~C.~E., {van de Ven}, G., {Verolme}, E.~K., {Cappellari},
  M., \& {de Zeeuw}, P.~T. 2008, \mnras, 385, 647

\bibitem[{{Vasiliev}(2019)}]{Vasiliev2019}
{Vasiliev}, E. 2019, \mnras, 482, 1525

\bibitem[{{Vogelsberger} {et~al.}(2014{\natexlab{a}}){Vogelsberger}, {Genel},
  {Springel}, {Torrey}, {Sijacki}, {Xu}, {Snyder}, {Bird}, {Nelson}, \&
  {Hernquist}}]{Vogelsberger2014b}
{Vogelsberger}, M., {Genel}, S., {Springel}, V., {et~al.} 2014{\natexlab{a}},
  \nat, 509, 177

\bibitem[{{Vogelsberger} {et~al.}(2014{\natexlab{b}}){Vogelsberger}, {Genel},
  {Springel}, {Torrey}, {Sijacki}, {Xu}, {Snyder}, {Nelson}, \&
  {Hernquist}}]{Vogelsberger2014}
{Vogelsberger}, M., {Genel}, S., {Springel}, V., {et~al.} 2014{\natexlab{b}},
  \mnras, 444, 1518

\bibitem[{{Vogelsberger} {et~al.}(2020){Vogelsberger}, {Marinacci}, {Torrey},
  \& {Puchwein}}]{Vogelsberger2020}
{Vogelsberger}, M., {Marinacci}, F., {Torrey}, P., \& {Puchwein}, E. 2020,
  Nature Reviews Physics, 2, 42

\bibitem[{{Walcher} {et~al.}(2014){Walcher}, {Wisotzki}, {Bekerait{\'e}},
  {Husemann}, {Iglesias-P{\'a}ramo}, {Backsmann}, {Barrera Ballesteros},
  {Catal{\'a}n-Torrecilla}, {Cortijo}, {del Olmo}, {Garcia Lorenzo},
  {Falc{\'o}n-Barroso}, {Jilkova}, {Kalinova}, {Mast}, {Marino},
  {M{\'e}ndez-Abreu}, {Pasquali}, {S{\'a}nchez}, {Trager}, {Zibetti},
  {Aguerri}, {Alves}, {Bland-Hawthorn}, {Boselli}, {Castillo Morales}, {Cid
  Fernandes}, {Flores}, {Galbany}, {Gallazzi}, {Garc{\'{\i}}a-Benito}, {Gil de
  Paz}, {Gonz{\'a}lez-Delgado}, {Jahnke}, {Jungwiert}, {Kehrig}, {Lyubenova},
  {M{\'a}rquez Perez}, {Masegosa}, {Monreal Ibero}, {P{\'e}rez}, {Quirrenbach},
  {Rosales-Ortega}, {Roth}, {Sanchez-Blazquez}, {Spekkens}, {Tundo}, {van de
  Ven}, {Verheijen}, {Vilchez}, \& {Ziegler}}]{Walcher2014}
{Walcher}, C.~J., {Wisotzki}, L., {Bekerait{\'e}}, S., {et~al.} 2014, \aap,
  569, A1

\bibitem[{{Wang} {et~al.}(2020){Wang}, {Wang}, {Mo}, {van den Bosch}, \&
  {Yang}}]{Wang2020}
{Wang}, E., {Wang}, H., {Mo}, H., {van den Bosch}, F.~C., \& {Yang}, X. 2020,
  \apj, 889, 37

\bibitem[{{Weinberger} {et~al.}(2017){Weinberger}, {Springel}, {Hernquist},
  {Pillepich}, {Marinacci}, {Pakmor}, {Nelson}, {Genel}, {Vogelsberger},
  {Naiman}, \& {Torrey}}]{Weinberger2017}
{Weinberger}, R., {Springel}, V., {Hernquist}, L., {et~al.} 2017, \mnras, 465,
  3291

\bibitem[{{Weinzirl} {et~al.}(2009){Weinzirl}, {Jogee}, {Khochfar}, {Burkert},
  \& {Kormendy}}]{Weinzirl2009}
{Weinzirl}, T., {Jogee}, S., {Khochfar}, S., {Burkert}, A., \& {Kormendy}, J.
  2009, \apj, 696, 411

\bibitem[{{Xu} {et~al.}(2019){Xu}, {Zhu}, {Grand}, {Springel}, {Mao}, {van de
  Ven}, {Lu}, {Wang}, {Pillepich}, {Genel}, {Nelson}, {Rodriguez-Gomez},
  {Pakmor}, {Weinberger}, {Marinacci}, {Vogelsberger}, {Torrey}, {Naiman}, \&
  {Hernquist}}]{Xu2019}
{Xu}, D., {Zhu}, L., {Grand}, R., {et~al.} 2019, \mnras, 489, 842

\bibitem[{{Zanisi} {et~al.}(2021){Zanisi}, {Huertas-Company}, {Lanusse},
  {Bottrell}, {Pillepich}, {Nelson}, {Rodriguez-Gomez}, {Shankar}, {Hernquist},
  {Dekel}, {Margalef-Bentabol}, {Vogelsberger}, \& {Primack}}]{Zanisi2021}
{Zanisi}, L., {Huertas-Company}, M., {Lanusse}, F., {et~al.} 2021, \mnras, 501,
  4359

\bibitem[{{Zhu} {et~al.}(2020){Zhu}, {van de Ven}, {Leaman}, {Grand },
  {Falc{\'o}n-Barroso}, {Jethwa}, {Watkins}, {Mao}, {Poci}, {McDermid}, \&
  {Nelson}}]{Zhu2020}
{Zhu}, L., {van de Ven}, G., {Leaman}, R., {et~al.} 2020, \mnras, 496, 1579

\bibitem[{{Zhu} {et~al.}(2018{\natexlab{a}}){Zhu}, {van de Ven},
  {M{\'e}ndez-Abreu}, \& {Obreja}}]{zhu2018c}
{Zhu}, L., {van de Ven}, G., {M{\'e}ndez-Abreu}, J., \& {Obreja}, A.
  2018{\natexlab{a}}, \mnras, 479, 945

\bibitem[{{Zhu} {et~al.}(2018{\natexlab{b}}){Zhu}, {van de Ven}, {van den
  Bosch}, {Rix}, {Lyubenova}, {Falc{\'o}n-Barroso}, {Martig}, {Mao}, {Xu},
  {Jin}, {Obreja}, {Grand }, {Dutton}, {Macci{\`o}}, {G{\'o}mez}, {Walcher},
  {Garc{\'\i}a-Benito}, {Zibetti}, \& {S{\'a}nchez}}]{zhu2018b}
{Zhu}, L., {van de Ven}, G., {van den Bosch}, R., {et~al.} 2018{\natexlab{b}},
  Nature Astronomy, 2, 233

\bibitem[{{Zhu} {et~al.}(2018{\natexlab{c}}){Zhu}, {van den Bosch}, {van de
  Ven}, {Lyubenova}, {Falc{\'o}n-Barroso}, {Meidt}, {Martig}, {Shen}, {Li},
  {Yildirim}, {Walcher}, \& {Sanchez}}]{Zhu2018a}
{Zhu}, L., {van den Bosch}, R., {van de Ven}, G., {et~al.} 2018{\natexlab{c}},
  \mnras, 473, 3000

\bibitem[{{Zolotov} {et~al.}(2009){Zolotov}, {Willman}, {Brooks}, {Governato},
  {Brook}, {Hogg}, {Quinn}, \& {Stinson}}]{Zolotov2009}
{Zolotov}, A., {Willman}, B., {Brooks}, A.~M., {et~al.} 2009, \apj, 702, 1058

\end{thebibliography}

\appendix
\section{Comparison of time-averaged and phase-space-averaged stellar orbit distributions}
\begin{figure*}
\centering\includegraphics[width=16cm]{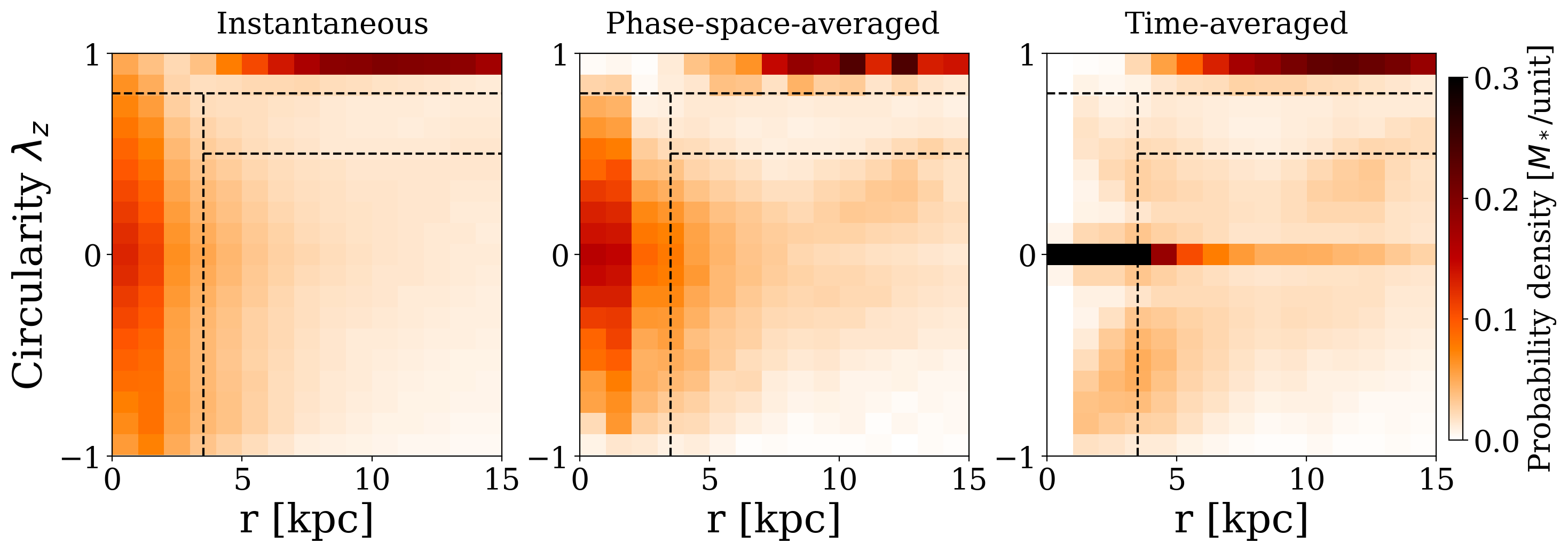}
\caption{Stellar orbit distribution, $p(r, \lambda_z)$, calculated in three different ways for an example galaxy from TNG50 at $z=0$. In the left, $r$ and $\lambda_z$ are calculated from the instantaneous positions and velocities of particles at $z=0$. In the middle, $r$ and $\lambda_z$ are phase-space-averaged. In the right, $r$ and $\lambda_z$ are time-averaged by integrating the orbits in the frozen potential. Most of the orbits in the bulge and in the hot inner stellar halo are on box orbits, and particles on the box orbits span a wide range of $\lambda_z$ but with time-averaged values of zero. By doing a phase-space averaging, we can narrow down the $\lambda_z$ distribution of particles on box orbits but cannot shrink them to zero. }
\label{fig:rlz_3}
\end{figure*}

In Fig.~\ref{fig:rlz_3} we show the stellar orbit distribution, $p(r, \lambda_z)$, calculated in three different ways for an example galaxy, TNG50 468590: (1) from the instantaneous positions and velocities of particles at $z=0$, (2) from phase-space-averaged values, and (3) from values that are time-averaged via the integration of the orbits in the frozen potential.
The hot inner stellar halo fractions calculated from the three panels are $21.5\%$, $24.9\%$,  and $24.5\%$, respectively, and the last two agree well with each other. This suggests that, for the purposes of the analysis here, it is sufficient to characterize galaxies in the $(r, \lambda_z)$ plane via phase-space averaging.

\section{Distribution of $r|_{\rm max(CDF deviation)}$}
\label{s:rcdf_9}
\begin{figure}
\centering\includegraphics[width=9cm]{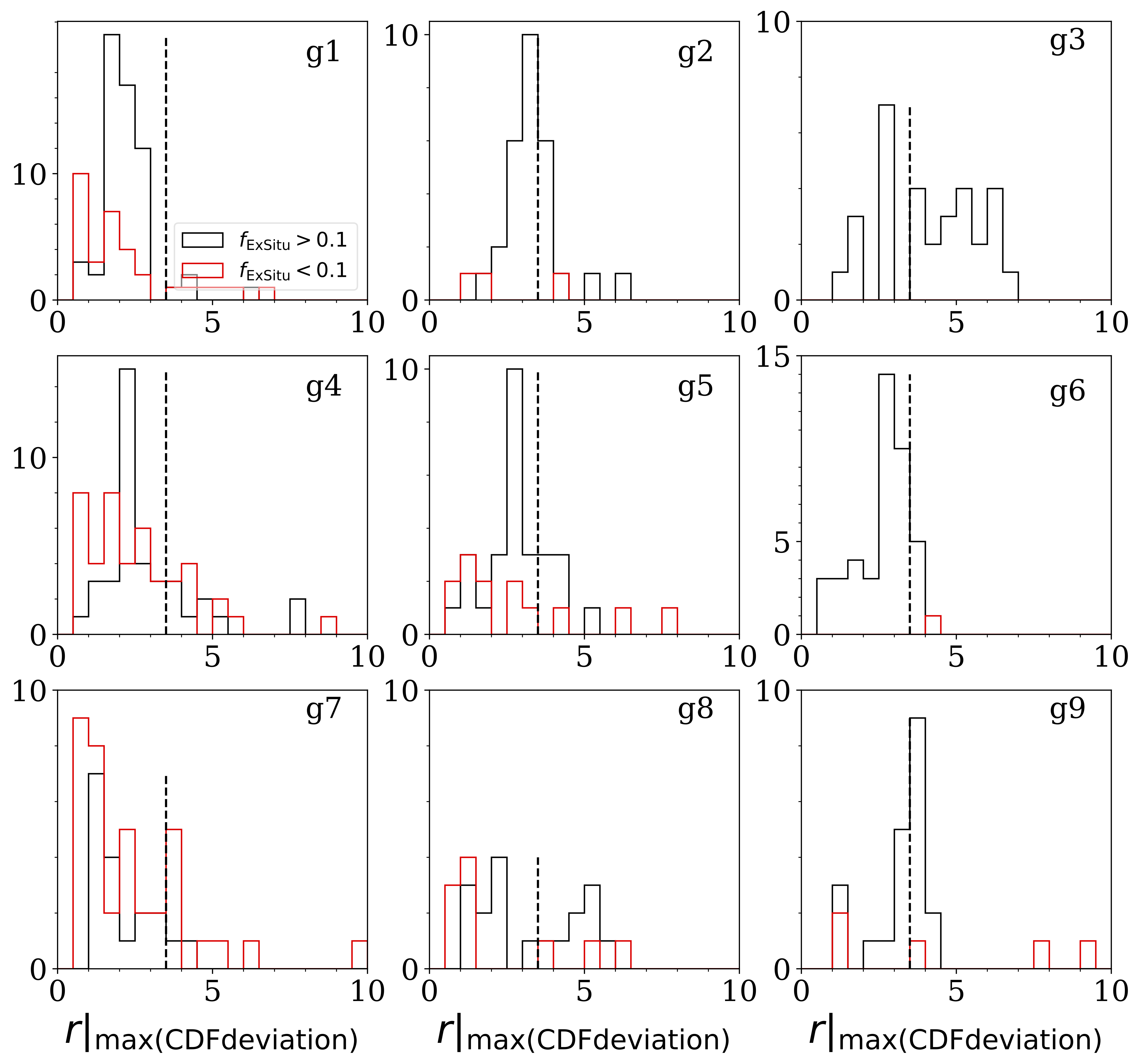}
\caption{Distribution of $r|_{\rm max(CDF deviation)}$, defined as the radius where ${\rm CDF}_{\rm hot} - {\rm CDF}_{\rm hot, control}$ reaches the maximum, for TNG50 galaxies in g1-g9. The  black histogram represents galaxies with higher ex situ fractions ($f_{\rm ExSitu} > 0.1$), and the red histogram galaxies with low ex situ fraction ($f_{\rm ExSitu} < 0.1$). }
\label{fig:rcdf_9}
\end{figure}

We can construct the deviation of ${\rm CDF}_{\rm hot}$ from a control model, ${\rm CDF}_{\rm hot, control}$, for all TNG50 galaxies from subsamples g1-g9. In each group, ${\rm CDF}_{\rm hot, control}$ is constructed by galaxies with $f_{\rm ExSitu}<0.05+\min(f_{\rm ExSitu})$ in that group, in a similar way as shown in Fig.~\ref{fig:rcut} for g6.
As described in the main text of the paper, we define the radius where ${\rm CDF}_{\rm hot}$ has the maximum deviation compared to the control model as $r|_{\rm max(CDF deviation)}$, which reflects the transition from ``bulge'' to the merger-induced hot inner stellar halo for galaxies with relatively high ex situ fractions. 
In Fig.~\ref{fig:rcdf_9} we show the distribution of $r|_{\rm max(CDF deviation)}$ for TNG50 galaxies.

For galaxies with $f_{\rm ExSitu} > 0.1$, $r|_{\rm max(CDF deviation)}$ has a narrow distribution around $3.5$ kpc: this is the case for galaxies in subsamples g2, g5, g6, g9, and, with larger variations, g3. The $r|_{\rm max(CDF deviation)}$ is smaller and distributed around 2.5 kpc for galaxies in g1 and g4 and around 2 kpc in g7.  
For galaxies with $f_{\rm ExSitu} < 0.1$, $r|_{\rm max(CDF deviation)}$ has a wide distribution in all groups of galaxies, and it cannot reflect the real bulge size. In these galaxies, the variation in the density distribution of hot orbits is dominated by the intrinsic variation in the bulge and is not induced by mergers.

\section{Correlation between the total stellar mass and accreted satellite mass}
\label{SS:mstar}
\begin{figure}
\centering
\includegraphics[width=8cm]{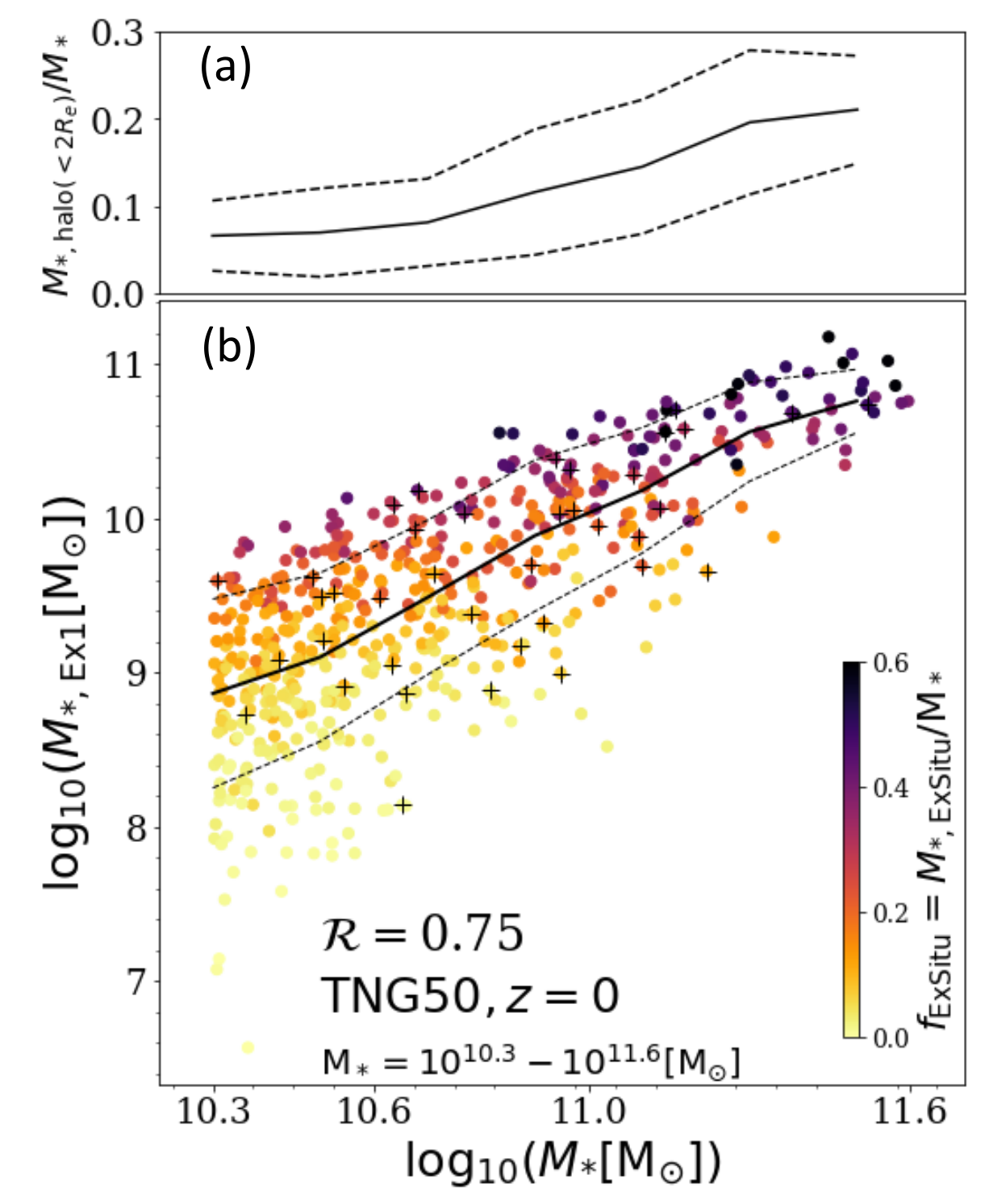}
\caption{Correlation between a galaxy's stellar mass, \Mstar, and the stellar mass accreted from the most massive merger according to TNG50. Panel (a): Mass fraction of the hot inner stellar halo, \Mshalo, compared to the total stellar mass, \Mstar, as a function of \Mstar\,. The solid line is the average, and the dashed lines are $\pm1\sigma$ scatter. In this mass range, the hot inner stellar halo mass contributes only $\sim 5\%-30\%$ of the total stellar mass. Panel (b): Similar to the right panel of Fig.~\ref{fig:Macc3_tng50_1}, but using the total stellar mass, $M_*$, instead of the hot inner stellar halo mass along the x axis. Compared to Fig.~\ref{fig:Macc3_tng50_1}, the correlation becomes weaker with $\mathcal{R}=0.75$ and the $1\sigma$ scatter is about 0.1 dex larger in general. The ex situ fraction does not increase as gradually with increasing x as in Fig.~\ref{fig:Macc3_tng50_1}, and the correlation of \Mstar\, with $f_{\rm ExSitu}$ is also weaker than \Mshalo\,. }
\label{fig:mstar_4sims}
\end{figure}

As a reference point to the main result of this paper (Fig.~\ref{fig:Macc3_tng50_1}), we show the correlation of a galaxy's stellar mass, \Mstar, with the stellar mass accreted from the most massive mergers, $M_{*,\rm{Ex1}}$, for all TNG50 galaxies with $10^{10.3}\leq\Mstar\leq10^{11.6}$\,\Msun\, in the main panel of Fig.~\ref{fig:mstar_4sims}. In comparison to the correlation between \Mshalo\, and $M_{*,\rm{Ex1}}$ (Fig.~\ref{fig:Macc3_tng50_1}, right panel), the correlation with total galaxy mass is weaker: the correlation coefficient reads $\mathcal{R}=0.75$ and the $1\sigma$ scatter is about 0.1 dex larger in general. 
For completeness, in the upper panel of Fig.~\ref{fig:mstar_4sims} we also show the typical fraction of stellar mass contained within the hot inner stellar halos to the total stellar mass, according to our definition of hot inner stellar halo and to TNG50.

\end{document}